\documentclass[aps,pra,twocolumn,showpacs,superscriptaddress]{revtex4}
\usepackage{graphicx}
\usepackage{amsmath}
\usepackage{amssymb}
\usepackage{lscape}
\usepackage{multirow}
\usepackage{longtable}
\usepackage{comment}
\usepackage{esvect}

\newcommand*\oline[1]{
   \vbox{
     \hrule height 0.5pt                
     \kern0.2ex                        
     \hbox{
       \kern-0.4em                      
       \ifmmode#1\else\ensuremath{#1}\fi
       \kern0.em
     }
   }
}

\input{epsf}

\begin{document}

\title{Energetics and structural properties of two- and three-boson systems
in the presence of 1D spin-orbit coupling}

\author{Q. Guan}
\affiliation{Homer L. Dodge Department of Physics and Astronomy,
  The University of Oklahoma,
  440 W. Brooks Street,
  Norman,
Oklahoma 73019, USA}
\affiliation{Center for Quantum Research and Technology,
  The University of Oklahoma,
  440 W. Brooks Street,
  Norman,
Oklahoma 73019, USA}

\author{D. Blume}
\affiliation{Homer L. Dodge Department of Physics and Astronomy,
  The University of Oklahoma,
  440 W. Brooks Street,
  Norman,
Oklahoma 73019, USA}
\affiliation{Center for Quantum Research and Technology,
  The University of Oklahoma,
  440 W. Brooks Street,
  Norman,
Oklahoma 73019, USA}

\date{\today}

\begin{abstract}
It was 
shown recently that the discrete scaling symmetry,
which underlies the Efimov effect in the three identical boson
system with two-body short-range interactions, survives when 
single-particle 1D spin-orbit coupling terms are added
to the Hamiltonian. 
Each three-body energy level in the ordinary Efimov scenario turns into an energy manifold 
that contains four energy levels
in the presence of 1D spin-orbit coupling
(equal mixture of Rashba-Dresselhaus coupling).
This work provides a detailed 
characterization of the energy levels in these manifolds.
The two-boson energies, 
which enter into the three-boson scattering threshold,
are analyzed in detail.
Moreover, the structural properties, e.g., momentum distributions
of the two- and three-boson systems,
are analyzed for various parameter combinations.
\end{abstract}

\maketitle

\section{Introduction}
\label{sec1}

The three-boson system with two-body short-range interactions
has captured physicists' attention ever since Efimov's
bizarre and counterintuitive prediction
that the strenthening of the two-body binding leads,
in a certain parameter regime, to a weakening of the 
three-body binding~\cite{efimov70, efimov71, efimov72, efimov73}. 
It is now well understood that this behavior is
linked to the existence of a discrete scaling symmetry and, 
associated with this symmetry, a limit cycle~\cite{braaten2006}.
The discrete scaling symmetry has been probed in atom-loss measurements
in cold, non-degenerate cold atom systems~\cite{grimm2006, Zaccanti, ferlaino2010, ferlaino2011, grimm2011, huangPRL}. 
Just a few years ago,
the quantum mechanical density of the helium  Efimov trimer system
was measured in a molecular beam experiment that allows for
size selection using matter wave diffraction and imaging via
Coulomb explosion~\cite{efimov_helium}. 
In a different experimental set-up,
radio-frequency spectroscopy was used to probe the energy spectrum~\cite{efimov_radio1, efimov_radio2}.

Ever since Efimov's prediction in the early 70s,
there has been a quest to extend the Efimov scenario beyond the
three identical boson paradigm.
In fact, motivated by possible applications to nuclear
systems, Efimov himself considered various extensions to
three-particle systems with different masses and spin
degrees of freedom~\cite{efimov73}.
Possible extensions to the four- and higher-body sector have
captured scientists' imagination and challenged our analytical
and numerical toolbox for the past fourty-plus years~\cite{GreenwoodPRD1973, SadhanPRD1981, PlatterPRA2004, StecherNatPhys2009, StecherJPhysB2010, NicholsonPRL2012, GattobigioPRA2014, YanPRA2015, naidon2016, ChrisRMP2017}.
More recently, possible imprints of three-body Efimov
physics on many-body systems have been investigated~\cite{BruunPRL2015, CuiPRL2017, TaylorPRL2017, DIncaoPRL2018, ParishPRX2018, CuiPRA2019}.
This work explores an extension of Efimov's scenario
along a different line, namely, it
considers the three-boson Efimov scenario in the
presence of single-particle forces. 
The most frequently considered single-particle force in the literature is an external
confinement, which reduces the ``position space'' available to the
three-particle system~\cite{JonsellPRL2002, BlumeRep, CastinPRL2006, KokkelmansFewBody2011, TolleJPhysG2013, ParishPRX2014, KohlerFoundPhysics2016}. 
Our work, in contrast, considers 
1D spin-orbit coupling terms, which modify the single-particle dispersion.
The impacts of other spin-orbit coupling schemes on Efimov trimer have been considered in the literature~\cite{shi2014, cui2014, shi2015}.

In a recent work~\cite{GuanPRX2018}, 
it was demonstrated that Efimov's
discrete scaling law persists
in the presence of 1D spin-orbit coupling
in an enlarged parameter space that
includes not only the two-body $s$-wave scattering length, 
but also the parameters that characterize the 1D spin-orbit coupling terms. 
The discrete scaling law tells us that
once we know the shape of four energy surfaces
in a five-dimensional parameter space,
we can predict all other energy surfaces in the five-dimensional parameter space.
This is similar to the ``normal'' Efimov scenario
where, once we know one energy curve in the energy-scattering length plane,
all other energy curves in this plane are determined by the
radial scaling law~\cite{braaten2006, naidon2016}. 
In this work, 
we map out the shape of the three-boson energy surfaces
in the presence of 1D spin-orbit coupling 
in a subspace of the full parameter space.    
Particular emphasis is given to the near-threshold behavior 
and its dependence on the generalized total momentum
(center-of-mass quasi-momentum).
The determination of the three-atom threshold requires the two-body 
energies as input. 
Because of this,
the two-boson system is investigated in detail.
In contrast to the extensively-studied two-fermion systems with various types of 
spin-orbit coupling
~\cite{bound_shenoy, crossover_shenoy, rashbon_shenoy, ShenoyPRA2013, xiaoling, PuPRA2013, zhenhua2012, zhenhua, xiaoling2017, qingze},
comparatively works on the two-boson systems with 2D and 3D spin-orbit coupling are found in the literature~\cite{qingze, wang2015, luo_boson, li_boson, xu_boson, YinPRA2018}. 

This work also considers structural properties of 
extremely weakly-bound two- and three-boson
systems in the presence of 1D spin-orbit coupling terms.
The 
momentum distributions 
of weakly-bound eigenstates help us understand
the effects of the 1D spin-orbit coupling terms on 
the binding energy.
In certain parameter regimes, 
the momentum distributions 
of weakly-bound states
and those of 
the lowest scattering threshold
are strongly correlated. 
Signatures of these correlations may be observable
in dedicated cold atom time-of-flight experiments.
Moreover, the generalized momentum and
the mechanical momentum in the lab frame are discussed 
for states with the strongest binding and 
for states with the lowest total energy.

The remainder of this paper is organized as follows.
Section~\ref{sec_system}
introduces the theoretical background. 
Sections~\ref{sec_twobody} and \ref{sec_threebody}
present our results for the two- and three-boson systems, respectively.
The dependence of the binding energy and the total ground state energy 
on the $s$-wave scattering length and the spin-orbit coupling parameters is analyzed. 
Momentum distribution functions are also analyzed.
Finally, Sec.~\ref{sec_conclusion} presents our conclusions.

\section{Theoretical background}
\label{sec_system}

\subsection{System Hamiltonian}
We consider $N$ identical bosons with mass $m$ and
three-dimensional position vectors $\vv{r}_j$.
Each atom is treated as an effective spin-1/2 system with spin-orbit coupling.
Using the Pauli spin-1/2 operators $\widehat{\sigma}_{j,x}$,
$\widehat{\sigma}_{j,y}$, and $\widehat{\sigma}_{j,z}$ for the $j$th particle,
the single-particle Hamiltonian $\widehat{h}_j$ reads~\cite{SpielmanNature2011}
\begin{eqnarray}
\label{h_single}
\widehat{h}_j = \frac{\widehat{\vv{p}}^2_j}{2m} \widehat{I}_j + 
\frac{\hbar k_{\text{so}}}{m} \widehat{p}_{j,z}
\widehat{\sigma}_{j,z}
+\frac{\Omega}{2} \widehat{\sigma}_{j,x} + \frac{\delta}{2} \widehat{\sigma}_{j,z},
\end{eqnarray}
where
$k_{\text{so}}$, $\Omega$, and $\delta$ are referred to as
spin-orbit coupling strength, Raman coupling, and detuning,
respectively.
The operator $\widehat{p}_{j,z}$ denotes the $z$-component
of the generalized momentum operator $\widehat{\vv{p}}_j$
of the $j$th atom ($\widehat{\vv{p}}_j$ contains the
components $\widehat{p}_{j,x}$, $\widehat{p}_{j,y}$, and $\widehat{p}_{j,z}$).
The quantity $\widehat{I}_j$ denotes the 2 by 2 identity operator in the spin space of the $j$th particle.
The system Hamiltonian $\widehat{H}_{N}$ for $N$ interacting particles reads
\begin{eqnarray}
\label{h_tot}
\widehat{H}_{N} = \sum_{j=1}^{N} \widehat{h}_j + \widehat{V}_{\text{int}},
\end{eqnarray} 
where 
the interaction term $\widehat{V}_{\text{int}}$ contains two-body interactions $V_{\text{2b}}$ and three-body interactions $V_{\text{3b}}$,
\begin{eqnarray}
\label{int}
  \widehat{V}_{\text{int}}= \left(\sum_{j=1,j<k}^N V_{\text{2b}}(r_{jk}) + \sum_{j=1,j<k,k<l}^N V_{\text{3b}}(r_{jkl})\right)\widehat{I}_{1,\cdots,N}.\\\nonumber
\end{eqnarray}
Here, we define the two-body relative distance $r_{jk}$ and 
three-body hyperradius $r_{jkl}$ as
\begin{eqnarray}
\label{two_body_relative}
r_{jk} = |\vv{r}_j-\vv{r}_k|
\end{eqnarray} 
and
\begin{eqnarray}
\label{three_body_relative}
r_{jkl} = (r_{jk}^2 + r_{jl}^2 + r_{kl}^2)^{1/2}.
\end{eqnarray} 
The quantity
$\widehat{I}_{j,\cdots,k}$ with $j \le k$
denotes the identity operator of the spin-space spanned by
particles $j$ through $k$. 

We use Jacobi coordinates~\cite{ECGbook, ECGrmp} to separate 
the center-of-mass degrees of freedom from the relative degrees of freedom.
Thus, the total system Hamiltonian $\widehat{H}_{N}$ in Eq.~\eqref{h_tot} can be rewritten as
\begin{eqnarray}
\label{h_tot_2}
\widehat{H}_{N} = \widehat{H}_{N, \text{com}} + \widehat{H}_{N, \text{rel}},
\end{eqnarray}
where the center-of-mass Hamiltonian $\widehat{H}_{N, \text{com}}$ and the 
relative Hamiltonian $\widehat{H}_{N, \text{rel}}$ take the forms 
\begin{eqnarray}
\label{h_com}
\widehat{H}_{N, \text{com}} = \frac{\widehat{\vv{q}}_N^2}{2\mu_N} \widehat{I}_{1,\cdots, N} + 
\frac{\hbar k_{\text{so}} N \widehat{q}_{N, z}}{\mu_N}
\widehat{\Sigma}_{N, z}
\end{eqnarray}
and
\begin{eqnarray}
\label{h_rel}
\widehat{H}_{N, \text{rel}}= 
\sum_{j=1}^{N-1} 
\frac{\widehat{\vv{q}}_j^2}{2 \mu_j} \widehat{I}_{1,\cdots,N} + 
\sum_{j=1}^{N-1}  \frac{\hbar k_{\text{so}}}{m} \widehat{q}_{j,z} 
\widehat{\Sigma}_{j,z}\\\nonumber
+ \frac{N \delta}{2}\widehat{\Sigma}_{N, z}
+ \frac{N \Omega}{2}\widehat{\Sigma}_{N, x}
+ \widehat{V}_{\text{int}}.
\end{eqnarray}
The quantity $\widehat{\vv{q}}_j$ ($j=1,\cdots,N$) denotes the
generalized $j$th Jacobi momentum operator and $\mu_j$
the associated Jacobi mass.
The transformation from the
generalized single-particle momentum
operators $\widehat{\vv{p}}_j$ to the generalized
Jacobi operators $\widehat{\vv{q}}_j$
is given by $(\widehat{\vv{p}}_1,\cdots,\widehat{\vv{p}}_N)^T
= U  (\widehat{\vv{q}}_1,\cdots,\widehat{\vv{q}}_N)^T$,
where 
the matrix $U$ is given by~\cite{ECGbook}
\begin{align}
U=
\begin{pmatrix}
1                 &    -1             & 0  & \cdots & 0\\
\frac{1}{2} & \frac{1}{2} & -1 & \cdots & 0\\
\vdots            & \                 &\   & \      & \vdots\\
\frac{1}{N-1} & \frac{1}{N-1} & \cdots & \cdots & -1\\
\frac{1}{N} & \frac{1}{N} & \cdots & \cdots & \frac{1}{N} \\
\end{pmatrix}.
\end{align}
The transformation from the single-particle
position vectors $\vv{r}_j$ to the Jacobi vectors
$\vv{\rho}_j$ (to be used below)
and from the single spin operators $\widehat{\vv{\sigma}}_{j}$
to the Jacobi spin operators $\widehat{\vv{\Sigma}}_{j}$ proceeds analogously
(Appendix A of Ref.~\cite{GuanPRX2018} provides explicit expressions for $\widehat{\Sigma}_{1,z}$ and $\widehat{\Sigma}_{2,z}$).

We identify the first term on the right hand side of Eq.~\eqref{h_com} 
as the kinetic energy associated with the center of mass degrees of freedom.
Since the total generalized momentum operator $\widehat{\vv{q}}_{N}$ is a conserved quantity~\cite{guan_thesis}, 
the eigenenergies and eigenstates of $\widehat{H}_N$ can be obtained by considering each 
fixed $\vv{q}_{N}$ subspace separately. 
Here, $\vv{q}_N$ is the eigenvalue of
the operator $\widehat{\vv{q}}_N$.
Similar to the system without spin-orbit coupling,
the kinetic energy associated with the center of mass degrees of freedom
contributes a constant energy shift to the eigenenergy for each fixed $\vv{q}_N$
and does not impact the binding energy.
The second term on the right hand side of Eq.~\eqref{h_com}
is structurally similar to the first term in the second line of Eq.~\eqref{h_rel}.
This discussion motivates us to define a modified relative Hamiltonian $\widehat{\oline{H}}_{N,\text{rel}}$,
which combines the second term on the right hand side of Eq.~\eqref{h_com}
and the first term in the second line of Eq.~\eqref{h_rel},
\begin{eqnarray}
\label{h_rel_bar}
\widehat{\oline{H}}_{N, \text{rel}}= 
\widehat{\oline{H}}_{N, \text{rel,ni}}+\widehat{V}_{\text{int}},
\end{eqnarray}
where the non-interacting relative Hamiltonian $\widehat{\oline{H}}_{N, \text{rel,ni}}$ reads
\begin{eqnarray}
\label{h_rel_bar_ni}
\widehat{\oline{H}}_{N, \text{rel,ni}}= 
\sum_{j=1}^{N-1} 
\frac{\widehat{\vv{q}}_j^2}{2 \mu_j} I_{1,\cdots,N} + 
\sum_{j=1}^{N-1}  \frac{\hbar k_{\text{so}}}{m} \widehat{q}_{j,z} 
\widehat{\Sigma}_{j,z}\\\nonumber
+ \frac{N \tilde{\delta}}{2}\widehat{\Sigma}_{N, z}
+ \frac{N \Omega}{2}\widehat{\Sigma}_{N, x}.
\end{eqnarray}
The generalized detuning $\tilde{\delta}$,
which contains the true detuning $\delta$ and the center-of-mass momentum $\vv{q}_{N}$,
is defined through 
\begin{eqnarray}
\label{eq_deltatilde}
\frac{\tilde{\delta}}{2} =
\frac{\hbar k_{\text{so}}}{\mu_N} q_{N,z} + \frac{\delta}{2}.
\end{eqnarray}
The shift introduces a non-trivial dependence of the eigenenergy on the total generalized momentum $\vv{q}_N$.
No such dependence exists for the system without spin-orbit coupling.
Using Eqs.~\eqref{h_rel_bar} and~\eqref{h_rel_bar_ni},
the total $N$-particle Hamiltonian for fixed $\vv{q}_N$ is given by
\begin{eqnarray}
\label{total_H_fix_q}
\widehat{H}_{N}(\vv{q}_N)=\frac{\vv{q}_N^2}{2\mu_N}I_{1,\cdots,N} +\widehat{\oline{H}}_{N,\text{rel}}.
\end{eqnarray} 
Equations~\eqref{h_rel_bar} and \eqref{h_rel_bar_ni} can be viewed as the system Hamiltonian in the center-of-mass frame and
the generalized detuning $\tilde{\delta}$ can be interpreted as the effective detuning
that the system sees in the center-of-mass frame.
The center-of-mass momentum dependence of $\widehat{\oline{H}}_{N, \text{rel,ni}}$ in Eq.~\eqref{h_rel_bar_ni} is 
a direct consequence of the breaking of the Galilean invariance in spin-orbit coupled systems~\cite{hui_review}.

In this work, 
the $\tilde{\delta}$-dependence of the binding energy $\oline{E}_{N,\text{binding}}^{(n)}(\tilde{\delta})$, 
where the superscript ``$n$'' indicates the $n$th eigenstate, is presented for two and three identical bosons.
The binding energy is directly related to the dissociation properties of few-body systems. 
For example, 
weakly-bound states tell one the critical scattering lengths 
at which enhanced three-body losses are expected to occur in the presence of 1D spin-orbit coupling~\cite{GuanPRX2018}.
Once the critical generalized detuning $\tilde{\delta}_{\text{cr}}$ 
at which the binding energy $\oline{E}^{(n)}_{N,\text{binding}}(\tilde{\delta})$ is the largest has been determined, 
the $z$-component of the generalized total momentum of
the corresponding state is uniquely determined via Eq.~\eqref{eq_deltatilde}
for each fixed $\delta$.
The same conclusion was reached by Shenoy~\cite{ShenoyPRA2013} 
for two identical fermions with 1D spin-orbit coupling.

\subsection{Binding energy}
\label{binding_energy_1}

To determine the binding energies $\oline{E}_{N,\text{binding}}^{(n)}(\tilde{\delta})$, 
one needs to calculate the threshold energy 
$\oline{E}_{N, \text{rel}, \text{th}}(\tilde{\delta})$ and the eigenenergy $\oline{E}^{(n)}_{N, \text{rel}}(\tilde{\delta})$ 
of the Hamiltonian $\widehat{\oline{H}}_{N,\text{rel}}$ for a given generalized detuning $\tilde{\delta}$,
\begin{eqnarray}
\label{def_binding}
\oline{E}^{(n)}_{N, \text{binding}}(\tilde{\delta})=\max\left[\oline{E}_{N, \text{rel}, \text{th}}(\tilde{\delta})-\oline{E}^{(n)}_{N, \text{rel}}(\tilde{\delta}), 0\right].
\end{eqnarray}
Equation~\eqref{def_binding} implies that,
assuming a fixed bare detuning $\delta$,
the eigenenergy of a state 
with total generalized momentum $\vv{q}_{N}$
is referenced to the 
threshold energy for a state with the same $\vv{q}_{N}$.
This is consistent with the fact that the components of $\vv{q}_N$
can be interpreted as good quantum numbers.
The threshold energy $\oline{E}_{N, \text{rel}, \text{th}}(\tilde{\delta})$ of the $N$-particle system
is the lowest eigenenergy of a state for
which one or more particles are far away from the rest of the system 
such that the interactions between the ``far-away particles'' and the rest of the system vanish.
For the two-body system, e.g.,
the threshold energy $\oline{E}_{2, \text{rel}, \text{th}}(\tilde{\delta})$ is equal to the lowest eigenenergy of 
two non-interacting particles~\cite{guan_thesis},
\begin{eqnarray}
\label{E_th_2}
\oline{E}_{2, \text{rel}, \text{th}}(\tilde{\delta}) = \min_{\vv{q}_1} \left[ \oline{E}^{(0)}_{2, \text{rel}, \text{ni}}(\vv{q}_1, \tilde{\delta})\right].
\end{eqnarray}
Here, $\oline{E}^{(0)}_{2, \text{rel}, \text{ni}}(\vv{q}_1, \tilde{\delta})$ corresponds to the lowest relative dispersion relationship
of the non-interacting Hamiltonian $\widehat{\oline{H}}_{2, \text{rel,ni}}$ with generalized detuning $\tilde{\delta}$
and relative generalized momentum $\vv{q}_1$.
Since the relative generalized momentum $\vv{q}_1$ is not a conserved quantity for
the two-particle Hamiltonian $\widehat{\oline{H}}_{2,\text{rel}}$,
the threshold energy is obtained by minimizing $\oline{E}_{2, \text{rel}, \text{ni}}^{(0)}(\vv{q}_1,\tilde{\delta})$ 
with respect to $\vv{q}_1$.

For the three-body system,
the threshold energy $\oline{E}_{3, \text{rel}, \text{th}}(\tilde{\delta})$ 
corresponds to either the lowest eigenenergy of three non-interacting particles
or to the lowest eigenenergy of an atom-dimer state (details are given in Appendices C-E of Ref.~\cite{GuanPRX2018}).
Specifically, 
we define the three-atom threshold energy $\oline{E}^{\text{aaa}}_{3, \text{rel}, \text{th}}(\tilde{\delta})$ 
and the atom-dimer threshold energy $\oline{E}^{\text{ad}}_{3, \text{rel}, \text{th}}(\tilde{\delta})$ 
through
\begin{eqnarray}
\label{E_th_3atom}
\oline{E}_{3,\text{rel},\text{th}}^{\text{aaa}}(\tilde{\delta}) = \min_{\vv{q}_1, \vv{q}_{2}}\left[\oline{E}_{3,\text{rel},\text{ni}}^{(0)}(\vv{q}_1, \vv{q}_{2}, \tilde{\delta})\right]
\end{eqnarray}
and
\begin{eqnarray}
\label{E_th_ad}
\nonumber
\oline{E}_{3,\text{rel,th}}^{\text{ad}}(\tilde{\delta}) = \min_{\vv{q}_{2}}\Bigg[\oline{E}_{2, \text{rel}}^{(0)}\left(\tilde{\delta}+\frac{\hbar k_{\text{so}}q_{2,z}}{m}\right)\\
+\oline{E}_{2,\text{rel,ni}}^{(0)}\left(\vv{q}_2, \tilde{\delta}-\frac{2\hbar k_{\text{so}}q_{2,z}}{m}\right)-\frac{\vv{q}_2^2}{4m}\Bigg],
\end{eqnarray}
respectively.
Here, $\oline{E}_{3,\text{rel,ni}}^{(0)}(\vv{q}_1, \vv{q}_{2}, \tilde{\delta})$ corresponds to the 
lowest dispersion relationship
of the non-interacting relative Hamiltonian $\widehat{\oline{H}}_{3, \text{rel,ni}}$ 
with generalized detuning $\tilde{\delta}$
and generalized Jacobi momenta $\vv{q}_1$ and $\vv{q}_2$.
Equation~\eqref{E_th_ad} shows that 
it is necessary to fully map out the $\tilde{\delta}$-dependence of $\oline{E}^{(0)}_{2, \text{rel}}(\tilde{\delta})$
to obtain $\oline{E}_{3,\text{rel,th}}^{\text{ad}}(\tilde{\delta})$.
Putting this together, 
the three-body threshold energy $\oline{E}_{3, \text{rel}, \text{th}}(\tilde{\delta})$ is determined by
\begin{eqnarray}
\label{E_th_3}
\oline{E}_{3, \text{rel}, \text{th}}(\tilde{\delta}) = \min\left[\oline{E}_{3, \text{rel}, \text{th}}^{\text{aaa}}(\tilde{\delta}), \oline{E}_{3, \text{rel}, \text{th}}^{\text{ad}}(\tilde{\delta})\right].
\end{eqnarray}

\subsection{Total ground state energy}

While the binding energy is relevant in the few-body context, 
the ground state energy $E_{N}^{(0)}(\vv{q}_N)$ of $\widehat{H}_N(\vv{q}_N)$ is
relevant in the many-body context.
Due to the breaking of the Galilean invariance, 
it is non-trivial to find the critical generalized total momentum $\vv{q}_{N,\text{cr}}$ at which 
$E_{N}^{(0)}(\vv{q}_N)$ reaches its minimum value, i.e., the total ground state energy.
This is different from the corresponding system without spin-orbit coupling,
where $E_{N}^{(n)}(\vv{q}_N)$ reaches its minimum value for $\vv{q}_N=0$. 
According to the Hellman-Feynman theorem~\cite{FeymanPhysRev1939}, we have
\begin{align}
\label{feynman_hellman}
\nabla_{\vv{q}_N}E_{N}^{(0)}(\vv{q}_N)=
\Bigg(\frac{q_{N, x}}{\mu_N},
\frac{q_{N, y}}{\mu_N},\\\nonumber
\frac{q_{N, z}}{\mu_N}+\frac{\hbar k_{\text{so}}N}{\mu_N}\langle\Psi_{\vv{q}_N}^{(0)} |\widehat{\Sigma}_{N, z}|\Psi_{\vv{q}_N}^{(0)}\rangle\Bigg)^T,
\end{align}
where $|\Psi_{\vv{q}_N}^{(0)}\rangle$ denotes the ground state of $\widehat{H}_{N}(\vv{q}_N)$ 
with generalized total momentum $\vv{q}_N$.
Since the critical generalized total momentum $\vv{q}_{N,\text{cr}}$
is defined through
\begin{eqnarray}
\label{critical_q}
\nabla_{\vv{q}_N}E_{N}^{(0)}(\vv{q}_N)\Big|_{\vv{q}_N = \vv{q}_{N, \text{cr}}} = 0,
\end{eqnarray}
Eq.~\eqref{feynman_hellman} yields
\begin{eqnarray}
\label{critical_q_2}
\vv{q}_{N,\text{cr}}=
\left(0, 0, -\hbar k_{\text{so}}N\langle\Psi_{\vv{q}_{N,\text{cr}}}^{(0)} |\widehat{\Sigma}_{N, z}|\Psi_{\vv{q}_{N,\text{cr}}}^{(0)}\rangle\right)^T.
\end{eqnarray}

In cold atom experiments, 
synthetic 1D spin-orbit coupling has first been realized using a Raman laser scheme~\cite{SpielmanNature2011}.
The derivation of the effective low-energy cold atom Hamiltonian involves
going from a bare state basis (lab frame) to a dressed state basis (rotated frame)~\cite{SpielmanNature2011}.
As a consequence, 
the total mechanical momentum $\widehat{\vv{q}}_{N,\text{lab}}$ in the lab frame is related to the generalized momentum $\widehat{\vv{q}}_{N}$ in the rotated frame via
\begin{eqnarray}
\label{lab_to_rotate}
\widehat{\vv{q}}_{N,\text{lab}} = \left(\widehat{q}_{N,x}, \widehat{q}_{N,y}, \widehat{q}_{N,z} + \hbar k_{\text{so}}N\widehat{\Sigma}_{N, z}\right)^T.
\end{eqnarray}
``Sandwiching'' Eq.~\eqref{lab_to_rotate} with the state $\Psi_{\vv{q}_{N,\text{cr}}}^{(0)}$,
one obtains
\begin{eqnarray}
\label{lab_to_rotate_critical}
\nonumber
\vv{q}_{N,\text{lab,cr}} = \Big(q_{N,x,\text{cr}}, q_{N,y,\text{cr}}, q_{N,z,\text{cr}} \\
+ \hbar k_{\text{so}}N\langle\Psi_{\vv{q}_{N,\text{cr}}}^{(0)}|\widehat{\Sigma}_{N, z}|\Psi_{\vv{q}_{N,\text{cr}}}^{(0)}\rangle\Big)^T,
\end{eqnarray}
where $q_{N,i,\text{cr}}$ $(i=x, y, z)$ denotes 
the $i$th component of the critical generalized total momentum $\vv{q}_{N,\text{cr}}$ 
and $\vv{q}_{N,\text{lab,cr}}$ the critical mechanical momentum in the lab frame.
Plugging Eq.~\eqref{critical_q_2} into Eq.~\eqref{lab_to_rotate_critical},
we find that $\Psi_{\vv{q}_{N,\text{cr}}}^{(0)}$ is characterized by a vanishing average total mechanical
momentum vector in the lab frame, 
regardless of the system parameters such as detuning and Raman coupling strength.
Said differently, 
the state corresponding to the lowest ground state energy $E^{(0)}_{N}(\vv{q}_{N,\text{cr}})$
among all ground state energies $E^{(0)}_{N}(\vv{q}_{N})$ 
has zero average total mechanical momentum in the lab frame.
This conclusion disagrees with the conclusions presented in Ref.~\cite{PuPRA2013}.

Given the ground state energy $\oline{E}^{(0)}_{N,\text{rel}}(\tilde{\delta})$ of $\widehat{\oline{H}}_{N, \text{rel}}$,
the ground state energy $E_{N}^{(0)}(\vv{q}_N)$ of $\widehat{H}_{N}(\vv{q}_N)$ is obtained through
\begin{eqnarray}
\label{E_tot_2}
E_{N}^{(0)}(\vv{q}_N) = \frac{\vv{q}_N^2}{2\mu_N} + \oline{E}^{(0)}_{N,\text{rel}}(\tilde{\delta}).
\end{eqnarray}
Equation~\eqref{E_tot_2} facilitates the process of mapping out the $\vv{q}_N$ dependence of $E_{N}^{(0)}(\vv{q}_N)$,
i.e.,
Eq.~\eqref{E_tot_2} 
serves as a ``hook'' that connects $E_{N}^{(0)}(\vv{q}_N)$ and $\oline{E}_{N,\text{rel}}^{(0)}(\tilde{\delta})$
for variable detunings $\delta$.
For example,
given $E_{N, \delta_1}^{(0)}(\vv{q}_N)$, 
$E_{N, \delta_2}^{(0)}(\vv{q}_N)$ is obtained via
\begin{eqnarray}
\label{eq_hook}
E_{N, \delta_2}^{(0)}(\vv{q}_{N}) = \frac{\vv{q}_{N}^2-\vv{Q}^2}{2\mu_N} + E_{N, \delta_1}^{(0)}(\vv{Q}), 
\end{eqnarray}   
where
\begin{eqnarray}
\label{Q}
\vv{Q} = \left(q_{N, x}, q_{N, y}, \frac{\left(\delta_2-\delta_1\right)\mu_N}{2\hbar k_{\text{so}}} + q_{N, z}\right).
\end{eqnarray}
In practice, 
we calculate the eigenenergies of $\widehat{\oline{H}}_{N,\text{rel}}$
as a function of the generalized detuning $\tilde{\delta}$
using the explicitly correlated Gaussian approach~\cite{ECGbook, ECGrmp, GuanPRX2018} 
(see Appendix B of Ref.~\cite{GuanPRX2018} for more details).
In a second step, we use Eqs.~\eqref{E_tot_2}-\eqref{Q} 
for a fixed bare detuning $\delta$ 
to obtain $E_{N}(\vv{q}_N)$ for various $\vv{q}_N$.

\subsection{Interaction potential and energy scales}

Throughout this work, we assume that the two-body
interaction
potential $V_{\text{2b}}(\vv{r}_{jk})$ is the same for all spin channels.
We use a Gaussian model potential with range $r_0$ and depth
$v_0$,
\begin{eqnarray}
\label{V2b}
V_{\text{2b}}(r_{jk}) = -v_0 \exp \left( -\frac{r_{jk}^2}{2r_0^2}
\right).
\end{eqnarray}
The depth $v_0$ ($v_0  \ge 0$) is adjusted to dial in the
desired two-body $s$-wave scattering length $a_s$. 
The $v_0$ values considered are such that the
potential $V_{\text{2b}}(r_{jk})$ supports at most one
two-body $s$-wave bound state.
The three-body interaction employed in this work also has a Gaussian form,
\begin{eqnarray}
\label{V3b}
V_{\text{3b}}(r_{jkl}) = V_0 \exp \left( -\frac{r_{jkl}^2}{2R_0^2}
\right).
\end{eqnarray}
The parameters $R_0$ and $V_0$ ($V_0\ge 0$) are used to tune the three-body parameter $\kappa_*$~\cite{GuanPRX2018, StecherJPhysB2010, GattobigioPRA2013, YanPRA2015},
\begin{eqnarray}
\label{kappa_star}
E_* = -\frac{\hbar^2 \kappa_*^2}{m},
\end{eqnarray}
where $E_*$ denotes the relative energy of the lowest universal
three-boson state at unitarity in the absence of spin-orbit coupling. 
The term ``universal three-boson state'' in this context
refers to a state described nearly perfectly by Efimov's zero-range theory~\cite{braaten2006, naidon2016}.
For the parameters considered in
this work, the most strongly bound three-boson state
is the lowest universal
three-boson state~\cite{GuanPRX2018}.
To guarantee that we are in the universal regime,
the ranges $r_0$ and $R_0$ in Eqs.~\eqref{V2b}-\eqref{V3b} 
are chosen to be much smaller than all other length scales
in the problem. 

In the following two sections,
we discuss 
the binding energies $\oline{E}^{(n)}_{N, \text{binding}}(\tilde{\delta})$,
the critical generalized detuning $\tilde{\delta}_{\text{cr}}$,
the ground state energy $E_{N}^{(0)}(\vv{q}_N)$ of $\widehat{H}_{N}(\vv{q}_N)$,
the critical total generalized momentum $\vv{q}_{N, \text{cr}}$,
and the total ground state energy $E_{N}^{(0)}(\vv{q}_{N, \text{cr}})$ for
bosonic systems with $N=2$ and $N=3$.
Since $E_{N}^{(0)}(\vv{q}_N)$ depends non-trivially only on the $z$-component of $\vv{q}_N$, 
we take $\vv{q}_N= (0, 0, q_{N,z})$ throughout
and use the notation $E_{N}^{(0)}(q_{N, z})$
when we discuss $E_{N}^{(0)}(\vv{q}_N)$.
Throughout this paper,
we use $E_{\text{so}}$ and $1/k_{\text{so}}$ as energy and length units,
\begin{eqnarray}
\label{E_so}
E_{\text{so}}=\frac{\hbar^2k_{\text{so}}^2}{2m}.
\end{eqnarray}

\section{Two-boson system} 
\label{sec_twobody}

\subsection{Binding energy}

\begin{figure}
\vspace*{0.cm}
\hspace*{0.cm}
\includegraphics[width=0.5\textwidth]{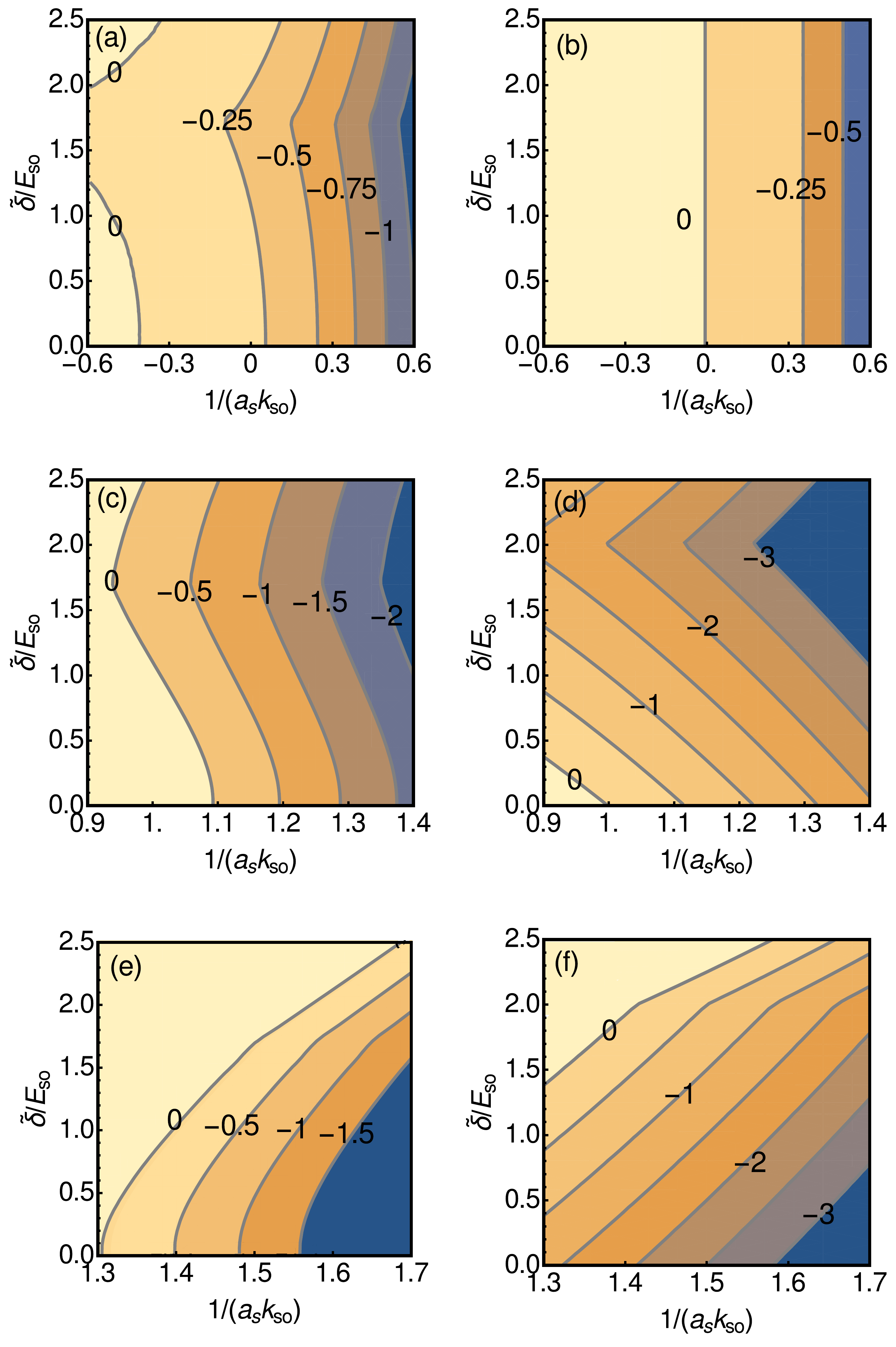}
\caption{(Color online)
The contours show the negative of the
two-boson binding energy, in units of $E_{\text{so}}$, 
as functions of $(a_s k_{\text{so}})^{-1}$ and $\tilde{\delta}/E_{\text{so}}$
for $\Omega=2E_{\text{so}}$ [panels~(a), (c), and (e)] and infinitesimally small $\Omega$ [panels~(b), (d), and (f)].
The panels in the first row, second row, and third row 
show the negative of the binding energy of the energetically
lowest-lying two-boson state $(n=0)$,
of the first excited two-boson state $(n=1)$,
and of the second excited two-boson state $(n=2)$.
Note the different ranges of the $x$-axis in panels~(a)-(f).
}
\label{fig_ebind_twobody}
\end{figure}

The two-body binding energies depend on the dimensionless parameters
$a_s k_{\text{so}}$, $\tilde{\delta}/E_{\text{so}}$, and $\Omega/E_{\text{so}}$.
For $\Omega=0$, i.e.,
in the absence of spin-orbit coupling, 
the total spin projection operator $\widehat{\Sigma}_{2,z}$ commutes with $\widehat{\oline{H}}_{2, \text{rel}}$.
This means that the associated $M_z$ projection quantum number is a good quantum number
and that the different spin channels are decoupled. 
Assuming that the eigenenergies for the case without spin-orbit coupling are known, 
the two-body eigenenergies for $\Omega=0$ can be obtained analytically 
(see Appendix~\ref{appendix_omegazero}).
In the zero-range limit, 
one finds that
the system supports up to three two-boson bound states.
The binding energy is in this $\Omega=0$ case measured with respect to the minimum of the 
respective non-interacting relative dispersion curve.
If we instead measured the
binding energy of the $\Omega=0$ system with respect to the absolute minimum
of the entire set of non-interacting relative dispersion curves
(which is how the two-atom threshold energy of the system with finite Raman coupling $\Omega$ is defined),
then we obtain the binding energies shown in Figs.~\ref{fig_ebind_twobody}(b),~\ref{fig_ebind_twobody}(d), and~\ref{fig_ebind_twobody}(f)
for the lowest two-boson state $(n=0)$, 
the first excited two-boson state $(n=1)$, 
and the second excited two-boson state $(n=2)$, respectively.
These binding energies,
calculated with respect to the incorrect threshold, 
can be interpreted as being those for infinitesimally small 
but finite $\Omega$.

For finite $\Omega$, 
$M_z$ is not a good quantum number any more and
the eigenstates are non-trivial superpositions of the four product spin states. 
Thus, the binding energies $\oline{E}_{2,\text{binding}}^{(n)}(\tilde{\delta})$ in the presence of spin-orbit coupling
have, in general, to be determined numerically.
As discussed in Sec~\ref{binding_energy_1}, 
the two-body binding energies $\oline{E}_{2,\text{binding}}^{(n)}(\tilde{\delta})$
are obtained by calculating the eigenenergies of the Hamiltonian $\widehat{\oline{H}}_{2,\text{rel}}$
for various generalized detunings $\tilde{\delta}$ 
and by measuring the energies relative to the lowest two-atom threshold energy 
$\oline{E}_{2,\text{rel,th}}(\tilde{\delta})$ with the same $\tilde{\delta}$.
Figures~\ref{fig_ebind_twobody}(a),~\ref{fig_ebind_twobody}(c), and~\ref{fig_ebind_twobody}(e)
show contour plots of
the negative of the binding energy $\oline{E}_{2,\text{binding}}^{(n)}(\tilde{\delta})$ 
for $n=0$, $n=1$, and $n=2$, respectively,
for $\Omega=2E_{\text{so}}$ as functions of $(a_s k_{\text{so}})^{-1}$ and $\tilde{\delta}/E_{\text{so}}$.
It is expected that the spin-orbit coupling has the
most pronounced effect on the eigenstates when the binding is weak.
To focus on the relatively weak binding regime, 
the range of the $(a_s k_{\text{so}})^{-1}$ values is different in
Figs.~\ref{fig_ebind_twobody}(a)-\ref{fig_ebind_twobody}(f).

Comparison of the left and right columns of Fig.~\ref{fig_ebind_twobody} shows that
the ``shapes'' of the contours for the first and second excited states for $\Omega=2E_{\text{so}}$ 
[Figs.~\ref{fig_ebind_twobody}(c) and~\ref{fig_ebind_twobody}(e)]
are quite similar to those for infinitesimally small $\Omega$
[Figs.~\ref{fig_ebind_twobody}(d) and~\ref{fig_ebind_twobody}(f)].
The binding energies of the first and second excited states are smaller 
for $\Omega=2E_{\text{so}}$ than for infinitesimally small $\Omega$.
For example, 
for $\tilde{\delta}=0$,
Fig.~\ref{fig_ebind_twobody}(c) shows that 
the first excited state starts to be bound at $(a_sk_{\text{so}})^{-1}\approx 1.1$
while Fig.~\ref{fig_ebind_twobody}(d) shows that it starts to be bound at $(a_sk_{\text{so}})^{-1}\approx 1$.
While the spin-orbit coupling ($\Omega=2E_{\text{so}}$) reduces the binding of the first excited and second excited states
compared to the case with infinitesimally small $\Omega$,
the binding of the ground state is enhanced by the finite spin-orbit coupling.

A key feature of
Fig.~\ref{fig_ebind_twobody}(a) is that the 
two-boson system supports a bound state on the
negative $s$-wave scattering length side.
The $s$-wave interacting system without spin-orbit coupling, in contrast,
does not support a bound state on the negative $s$-wave scattering length side 
[see Fig.~\ref{fig_ebind_twobody}(b)]. 
This implies that the spin-orbit coupling leads to an enhancement
of the binding of the two-boson ground state.
For $\Omega=2E_{\text{so}}$,
this enhancement is largest for a finite $\tilde{\delta}$, i.e.,
the critical generalized detuning 
is approximately equal to $1.7 E_{\text{so}}$ for all $a_sk_{\text{so}}$ included in Fig.~\ref{fig_ebind_twobody}(a).
For $\tilde{\delta}=1.7E_{\text{so}}$ and $\Omega=2E_{\text{so}}$, 
the smallest $(a_s k_{\text{so}})^{-1}$
value for which the two-boson system supports a bound state is equal to $-1.016$.
Assuming $(k_{\text{so}})^{-1}=4,000a_0$~\cite{SpielmanNature2011}, 
this corresponds to $a_s=-3,937a_0$.
This estimate shows that the enhancement of the two-body binding energy due
to the 1D spin-orbit coupling is sizable in the equal scattering lengths case considered in our work.
Although the experimentally more relevant unequal scattering lengths scenario
requires separate calculations,
the qualitative behavior is expected to be similar to that discussed here for the
identical scattering lengths case.

\begin{figure}
\vspace*{0.cm}
\hspace*{0.cm}
\includegraphics[width=0.35\textwidth]{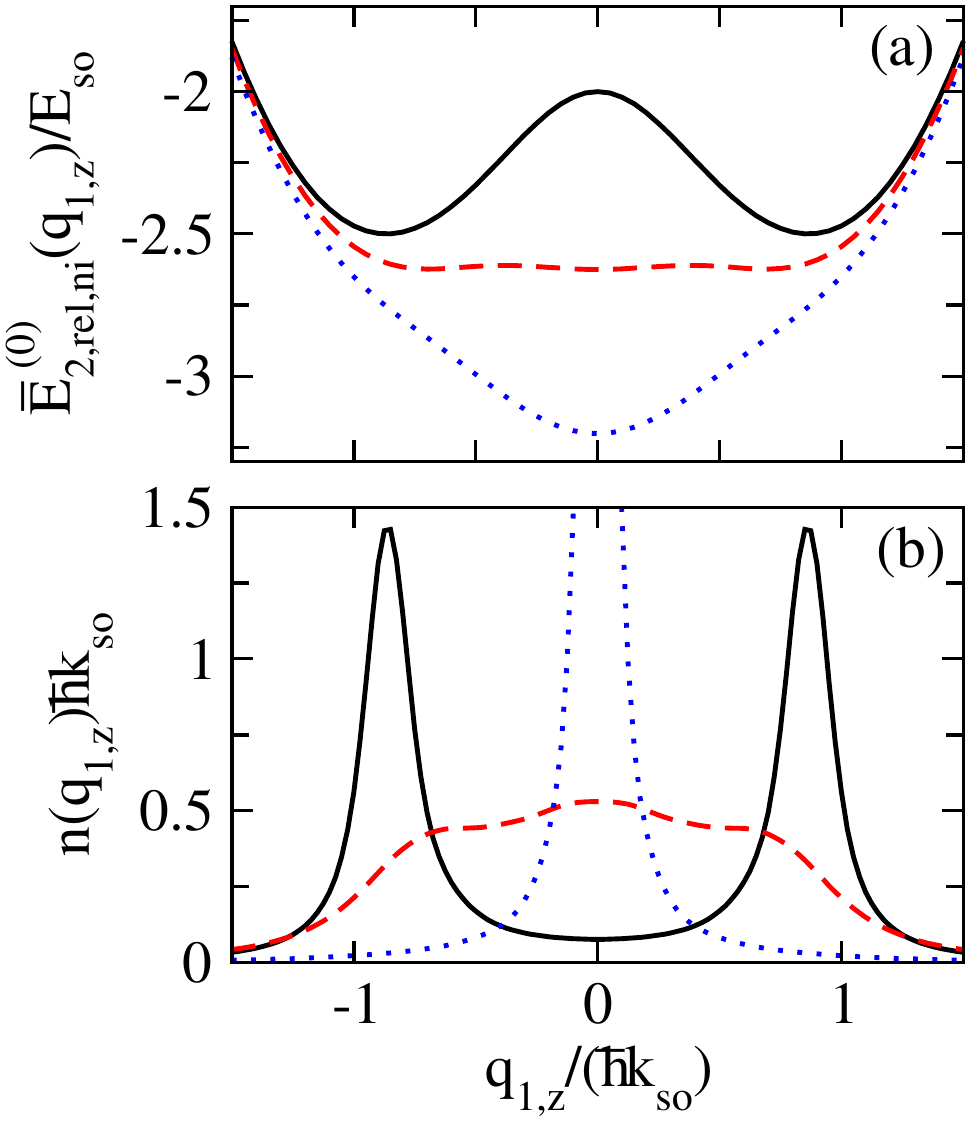}
\caption{(Color online)
Correlations between  
(a) the lowest non-interacting relative dispersion curve
and (b) the relative momentum distribution for the ground state
with $\Omega=2E_{\text{so}}$ and $(a_sk_{\text{so}})^{-1}=-0.321$.
For both panels,
the black solid, red dashed, and blue dotted lines correspond to 
$\tilde{\delta}=0$, $1.7E_{\text{so}}$, and $2.5E_{\text{so}}$, respectively.
}
\label{fig2}
\end{figure}

To gain a deeper understanding of
the shape of the contours in Figs.~\ref{fig_ebind_twobody}(a), 
it is helpful to analyze the lowest non-interacting relative dispersion relationship $\oline{E}^{(0)}_{2,\text{rel}, \text{ni}}(q_{1,z},\tilde{\delta})$ 
[see Eq.~\eqref{h_rel_bar_ni}].
Figure~\ref{fig2}(a) shows 
$\oline{E}^{(0)}_{2,\text{rel,ni}}(q_{1,z},\tilde{\delta})$ 
for three different generalized detunings 
$\tilde{\delta}$.
For $\tilde{\delta}=0$ 
[the black solid line in Fig.~\ref{fig2}(a)],
$\oline{E}^{(0)}_{2,\text{rel,ni}}(q_{1,z},\tilde{\delta})$ 
has two global minima at finite $q_{1,z}$.
For $0<\tilde{\delta}<1.697E_{\text{so}}$ (not shown),
a local minimum exists at $q_{1,z}=0$ in addition to the 
two global minima at finite $q_{1,z}$.
For $\tilde{\delta}=1.697E_{\text{so}}$ 
[the red dashed line in Fig.~\ref{fig2}(a)],
the minimum at $q_{1,z}=0$ is degenerate 
with the two minima at finite $q_{1,z}$;
thus, 
$\oline{E}^{(0)}_{2,\text{rel,ni}}(q_{1,z},\tilde{\delta})$ 
has three global minima.
For $\tilde{\delta}>1.697E_{\text{so}}$ 
[the blue dotted line in Fig.~\ref{fig2}(a)],
$\oline{E}^{(0)}_{2,\text{rel,ni}}(q_{1,z},\tilde{\delta})$ 
has one global minimum that is located at $q_{1,z}=0$.
For fixed 
$\tilde{\delta}$,
the minimum of 
$\oline{E}^{(0)}_{2,\text{rel,ni}}(q_{1,z},\tilde{\delta})$ 
yields the threshold energy 
$\oline{E}^{(0)}_{2,\text{rel,th}}(\tilde{\delta})$.
Thus, as $\tilde{\delta}$ increases, 
the degeneracy of the threshold energy
goes from two for $\tilde{\delta}<1.697E_{\text{so}}$ to three for $\tilde{\delta}=1.697E_{\text{so}}$, 
to one for $\tilde{\delta}>1.697E_{\text{so}}$. 
Figure~\ref{fig_ebind_twobody}(a) 
indicates that 
the generalized detuning at which 
the degeneracy of the threshold energy is maximal
is approximately equal to the critical detuning $\tilde{\delta}_{\text{cr}}$
at which the binding energy is largest.  
The same conclusion also holds in the three-boson system 
in the presence of 1D spin-orbit coupling 
(see Sec.~\ref{sec_threebody}).
 
In the weakly bound regime, 
the structural properties of the bound states are expected to reflect  
the behavior of the associated non-interacting relative dispersion curves. 
As an example,
we consider the relative momentum distribution $n(q_{1,z})$ along the $z$-direction 
for various generalized detunings.
The relative momentum distribution $n(q_{1,z})$ is defined through
\begin{eqnarray}
\label{two_body_momentum}
n(q_{1,z})=\sum_{\sigma}\int\Phi_{\text{rel},\sigma}^*(\vv{q}'_1)\delta(q_{1,z}'-q_{1,z})\Phi_{\text{rel},\sigma}(\vv{q}'_1)d\vv{q}'_{1},
\end{eqnarray}
where $\Phi_{\text{rel},\sigma}$ is the 
momentum space wave function associated with the spin component $\sigma$
($\sigma=|\uparrow\uparrow\rangle$, $|\uparrow\downarrow\rangle$, $|\downarrow\uparrow\rangle$, and $|\downarrow\downarrow\rangle$) 
of the eigenstate of the relative Hamiltonian $\oline{H}_{2,\text{rel}}$
and $\delta(q_{1,z}'-q_{1,z})$ is the Dirac delta function. 

Figure~\ref{fig2}(b) shows $n(q_{1,z})$ for the ground state 
for $\Omega=2E_{\text{so}}$ and $(a_sk_{\text{so}})^{-1}=-0.321$
[same as in Fig.~\ref{fig_ebind_twobody}(a)].
For $\tilde{\delta}=0$ [black solid line in Fig.~\ref{fig2}(b)],
$n(q_{1,z})$ has two distinct peaks located at finite $q_{1,z}$.
Due to the spin-momentum locking, 
the ground state is primarily a superposition of two spin contributions, namely, $|\uparrow\downarrow\rangle$
and $|\downarrow\uparrow\rangle$. 
For $\tilde{\delta}=1.7E_{\text{so}}$ [red dashed line in Fig.~\ref{fig2}(b)],
$n(q_{1,z})$ has three momentum peaks, 
two at finite $q_{1,z}$ and one at vanishing $q_{1,z}$.
In this case, 
the ground state has significant weights for the $|\uparrow\downarrow\rangle$, 
$|\downarrow\uparrow\rangle$, and $|\downarrow\downarrow\rangle$ components.
For $\tilde{\delta}=2.5E_{\text{so}}$ 
[blue dotted line in Fig.~\ref{fig2}(b)],
the ground state has one peak located at vanishing $q_{1,z}$
and consists predominantly of the $|\downarrow\downarrow\rangle$ spin component.
Comparing the curves in Figs.~\ref{fig2}(a) and~\ref{fig2}(b) 
for the same generalized detuning,
it can be seen that the $q_{1,z}$ values for which $n(q_{1,z})$ reaches a maximum
are correlated with those for which the corresponding 
lowest relative dispersion curve reaches a minimum.

\subsection{Total ground state energy}

\begin{figure}
\vspace*{0.2cm}
\hspace*{0.cm}
\includegraphics[width=0.48\textwidth]{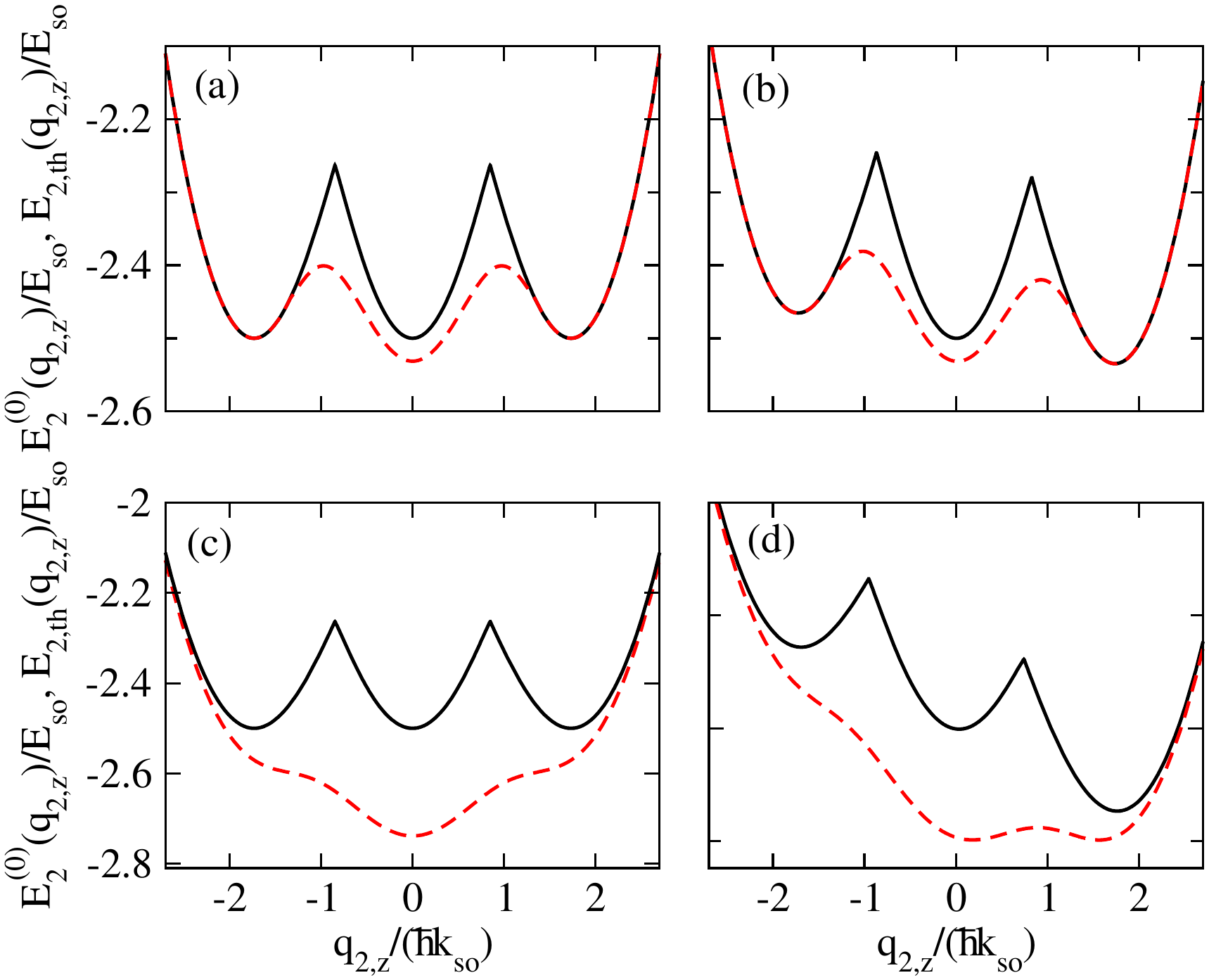}
\caption{(Color online)
The ground state energy $E_{2}^{(0)}(q_{2,z})$ of $\widehat{H}_{N}(q_{2,z})$ (red dashed lines) 
and the threshold energy $E_{2,\text{th}}(q_{2,z})$ (black solid lines) for $\Omega=2E_{\text{so}}$
and various $(a_sk_{\text{so}})^{-1}$ and $\delta/E_{\text{so}}$ combinations.
Panels (a) and (b) correspond to $(a_sk_{\text{so}})^{-1}=-0.321$.
Panels (c) and (d) correspond to $(a_sk_{\text{so}})^{-1}=0.0113$.
Panels (a) and (c) correspond to $\delta=0$.
Panel (b) and (d) correspond to $\delta=0.04E_{\text{so}}$ and $\delta=0.21E_{\text{so}}$, respectively.
}
\label{fig3}
\end{figure}

Given the ground state energy $\oline{E}^{(0)}_{2,\text{rel}}(\tilde{\delta})$ 
of the relative Hamiltonian $\widehat{\oline{H}}_{2,\text{rel}}$,
the ground state energy $E_{2}^{(0)}(q_{2,z})$ of the full Hamiltonian $\widehat{H}_{N}(q_{2,z})$ is obtained using Eq.~\eqref{E_tot_2}.
Dashed and solid lines in
Fig.~\ref{fig3} show the ground state energy $E_{2}^{(0)}(q_{2,z})$ 
and the threshold energy $E_{2,\text{th}}(q_{2,z})$, respectively, 
as a function of $q_{2,z}$
for $\Omega=2E_{\text{so}}$ and 
various $(a_sk_{\text{so}})^{-1}$ and $\delta/E_{\text{so}}$ combinations. 
Figures~\ref{fig3}(a) and~\ref{fig3}(b) cover the weakly bound regime
while Figs.~\ref{fig3}(c) and~\ref{fig3}(d) cover the more strongly bound regime. 
For $(a_sk_{\text{so}})^{-1}=-0.321$ and $\delta=0$ 
[Fig.~\ref{fig3}(a)], 
the threshold energy $E_{2,\text{th}}(q_{2,z})$ has three global minima, 
two at finite $q_{2,z}$ and one at $q_{2,z}=0$. 
In this case, 
the ground state energy $E_{2}^{(0)}(q_{2,z})$ has two local minima at finite $q_{2,z}$
and one global minimum at $q_{2,z}=0$.
The former minima correspond to scattering states and the latter minimum to a bound state. 
For a non-zero bare detuning $\delta$,
both $E_{2,\text{th}}(q_{2,z})$ and $E_{2}^{(0)}(q_{2,z})$ are ``tilted'' 
and asymmetric with respect to $q_{2,z}=0$.
For $(a_sk_{\text{so}})^{-1}=-0.321$ 
and $\delta=0.04E_{\text{so}}$ [Fig.~\ref{fig3}(b)],
the minimum of $E_{2}^{(0)}(q_{2,z})$ at $q_{2,z}=1.739\hbar k_{\text{so}}$ 
is degenerate with the minimum at $q_{2,z}=0$.
The minimum at $q_{2,z}=0$ corresponds to a bound state
while the minimum at $q_{2,z}=1.739\hbar k_{\text{so}}$
corresponds to a scattering state. 
For $(a_sk_{\text{so}})^{-1}=-0.321$ and $\delta>0.04E_{\text{so}}$ (not shown),
both $E_{2,\text{th}}(q_{2,z})$ and $E_{2}(q_{2,z})$
possess a global minimum at $q_{2,z}\approx 1.7\hbar k_{\text{so}}$.
In this case, 
the global minimum of $E_{2}(q_{2,z})$ corresponds to a scattering state.

Figures~\ref{fig3}(c) and~\ref{fig3}(d) show the results for $(a_sk_{\text{so}})^{-1}=0.0113$, 
i.e., the more strongly bound regime.
For $\delta=0$ [Fig.~\ref{fig3}(c)],
the ground state energy $E_{2}^{(0)}(q_{2,z})$ has one global minimum at $q_{2,z}=0$. 
For $\delta=0.21E_{\text{so}}$ [Fig.~\ref{fig3}(d)], in contrast,
the ground state energy $E_{2}^{(0)}(q_{2,z})$ has two degenerate global minima, 
one at $q_{2,z}\approx 0$ 
and the other at $q_{2,z}\approx 1.7\hbar k_{\text{so}}$. 
Compared to the case shown in Fig.~\ref{fig3}(b),
the minimum closest to $1.7\hbar k_{\text{so}}$ corresponds to a bound state 
instead of a scattering state.
For even larger $\delta$,
$E_{2}^{(0)}(q_{2,z})$ has a non-degenerate global minimum that corresponds to a bound state
at $q_{2,z}\approx 1.7\hbar k_{\text{so}}$ .

\begin{figure}
\vspace*{0.2cm}
\hspace*{0.cm}
\includegraphics[width=0.40\textwidth]{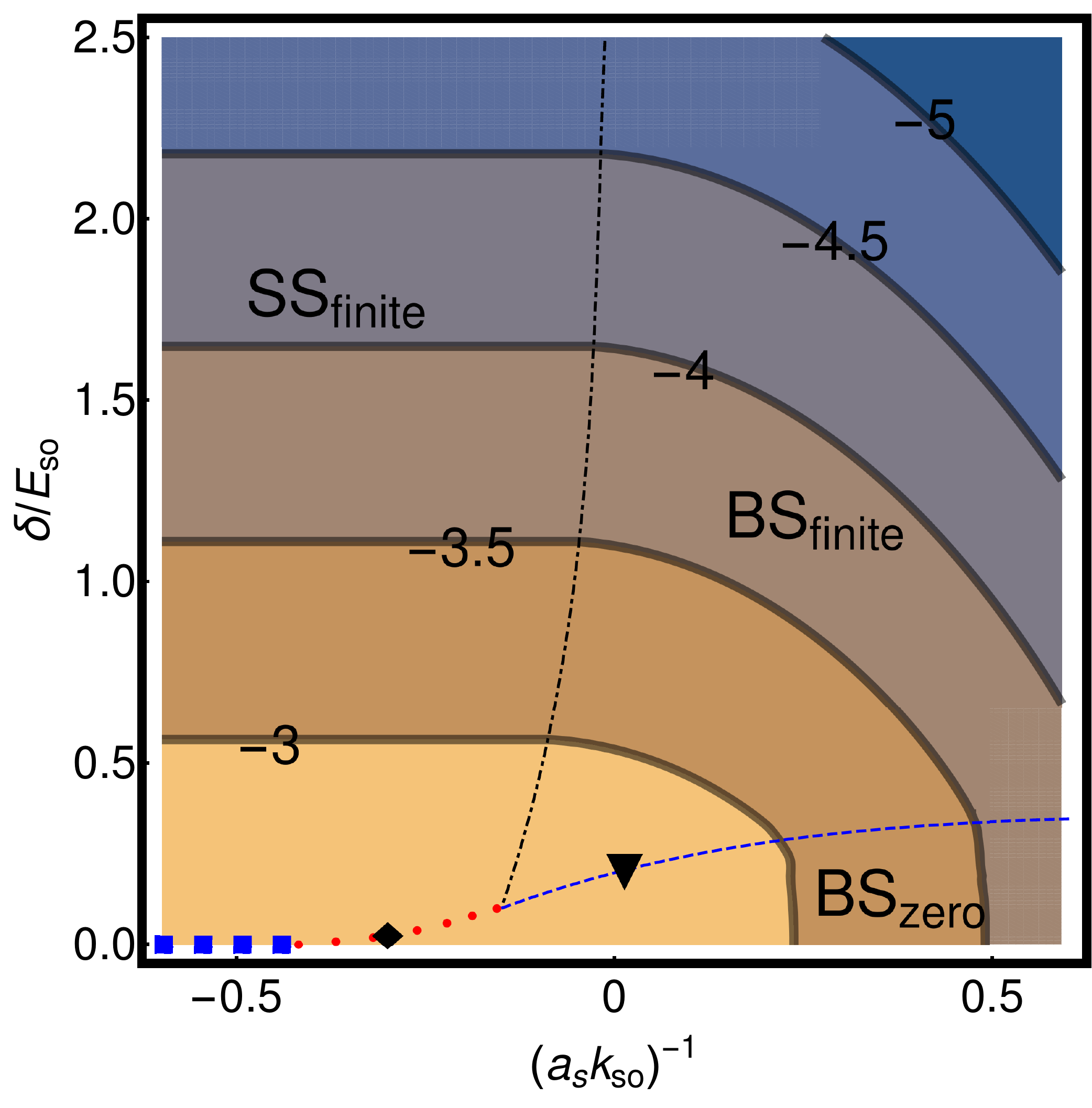}
\caption{(Color online)
``Phase diagram'' for the total ground state  
as functions of $(a_sk_{\text{so}})^{-1}$ and $\delta/E_{\text{so}}$ for $\Omega=2E_{\text{so}}$.
The contours show the total ground state energy $E_2^{(0)}(q_{2,z})/E_{\text{so}}$.
The black dot-dashed line, blue dashed line, red dotted line, and the blue squares separate the three phases $\text{SS}_{\text{finite}}$,
$\text{BS}_{\text{finite}}$, and $\text{BS}_{\text{zero}}$ from each other (see text for details).
The diamond and triangle mark the parameter combinations corresponding to Figs.~\ref{fig3}(b) and~\ref{fig3}(d), respectively.
}
\label{fig4}
\end{figure}

Figure~\ref{fig3} indicates 
that for each $(a_sk_{\text{so}})^{-1}$,
there exists a bare detuning $\delta$
for which the critical generalized total momentum $q_{2,z,\text{cr}}$ 
jumps from a value close to zero to a value 
close to $1.7\hbar k_{\text{so}}$.
Depending on the value of $(a_sk_{\text{so}})^{-1}$,
the global minimum of $E_{2}^{(0)}(q_{2,z})$
corresponds either to a scattering state or
to a bound state.
Thus, 
we can identify different ``phases'' for fixed $\Omega/E_{\text{so}}$, 
which categorize the total ground state. 
Figure~\ref{fig4} shows the ``phase diagram'' for 
the total ground state 
as functions of $(a_sk_{\text{so}})^{-1}$ and $\delta/E_{\text{so}}$
for $\Omega=2E_{\text{so}}$.
For this $\Omega/E_{\text{so}}$,
the total ground state falls in one of the following three phases.
$\text{SS}_{\text{finite}}$: The ground state is a scattering state with $q_{z,\text{cr},z}\ne 0$.
$\text{BS}_{\text{finite}}$: The ground state is a bound state with $q_{z,\text{cr},z}\approx 1.7\hbar k_{\text{so}}$.
$\text{BS}_{\text{zero}}$: The ground state is a bound state with $q_{z,\text{cr},z}\approx 0$.
The phases $\text{SS}_{\text{finite}}$ and $\text{BS}_{\text{finite}}$ have no analogy in the two-body system 
without spin-orbit coupling.

The region encircled by the blue squares, the red dotted line, 
the black dot-dashed line, and the upper and left edge of the figure
corresponds to the phase $\text{SS}_{\text{finite}}$.
The region encircled by the lower and right edge of the figure,
the blue dashed line, and the red dotted line
corresponds to the phase $\text{BS}_{\text{zero}}$. 
The region encircled by the blue dashed line, the right and upper edge of the figure, and the black dot-dashed line
corresponds to the phase $\text{BS}_{\text{finite}}$.
Along the red dotted line
[Fig.~\ref{fig3}(b) corresponds to such a situation],
the total ground state is two-fold degenerate:
one state corresponds to a scattering state with $q_{2,z}\approx 1.7\hbar k_{\text{so}}$ 
and the other to a bound state with $q_{2,z}\approx 0$.
Along the blue dashed line
[Fig.~\ref{fig3}(d) corresponds to such a situation],
the total ground state is two-fold degenerate:
both states correspond to bound states but with different $q_{2,z}$, 
one has $q_{2,z}\approx 1.7\hbar k_{\text{so}}$ 
and the other $q_{2,z}\approx 0$.
Along the blue squares, 
the total ground state is three-fold degenerate:
one state corresponds to a scattering state with $q_{2,z}=0$
and the other two to bound states with $q_{2,z}=\pm 1.739\hbar k_{\text{so}}$.  
Along the black dot-dashed line,
the total ground state is one-fold degenerate
and has a total momentum $q_{2,z}\approx 1.7\hbar k_{\text{so}}$.
For $\delta\rightarrow\infty$,
the black dot-dashed line in Fig.~\ref{fig4} is characterized by $(a_sk_{\text{so}})^{-1}=0$,
i.e., the role of $\Omega$ decreases with increasing $\delta$.

\section{Three-boson system}
\label{sec_threebody}

The three-boson properties depend on $a_sk_{\text{so}}$, $\tilde{\delta}/E_{\text{so}}$, $\Omega/E_{\text{so}}$, and $\kappa_*/k_{\text{so}}$.
The latter is the three-body parameter associated with Efimov physics. 
Throughout this section, we use $R_0/r_0=\sqrt{8}$ and fix the height of the three-body potential 
such that the lowest three-boson state at unitarity in the absence of spin-orbit coupling is characterized by $\kappa_*r_0= 0.0152$.
This state can be considered an Efimov state, i.e., upon variation of $a_s$
it follows Efimov's radial scaling law quite accurately. 
For example, the binding energy of the next excited state is 515.29 times smaller than the energy of the state we are considering at unitarity. 
This ratio is within $0.1\%$ of the scaling factor of Efimov's zero-range theory. 
Throughout this section, 
we fix $k_{\text{so}}$ (and correspondingly $E_{\text{so}}$)
and vary $\tilde{\delta}$ and $\Omega$.
Specifically, 
we choose $k_{\text{so}}$ such that it is comparable to $\kappa_*$,
$k_{\text{so}}=1.32\kappa_{*}$ and $E_{\text{so}}=-0.871E_*$.
Due to the close match of the energy scales, 
this parameter combination is expected to lead to significant modification of,
for example, the momentum distribution of the Efimov trimer near unitarity, i.e., near $|a_s|^{-1}\approx 0$.

\subsection{Binding energy}
\label{sec_fourb}

\begin{figure}
\vspace*{0.cm}
\hspace*{0.cm}
\includegraphics[width=0.5\textwidth]{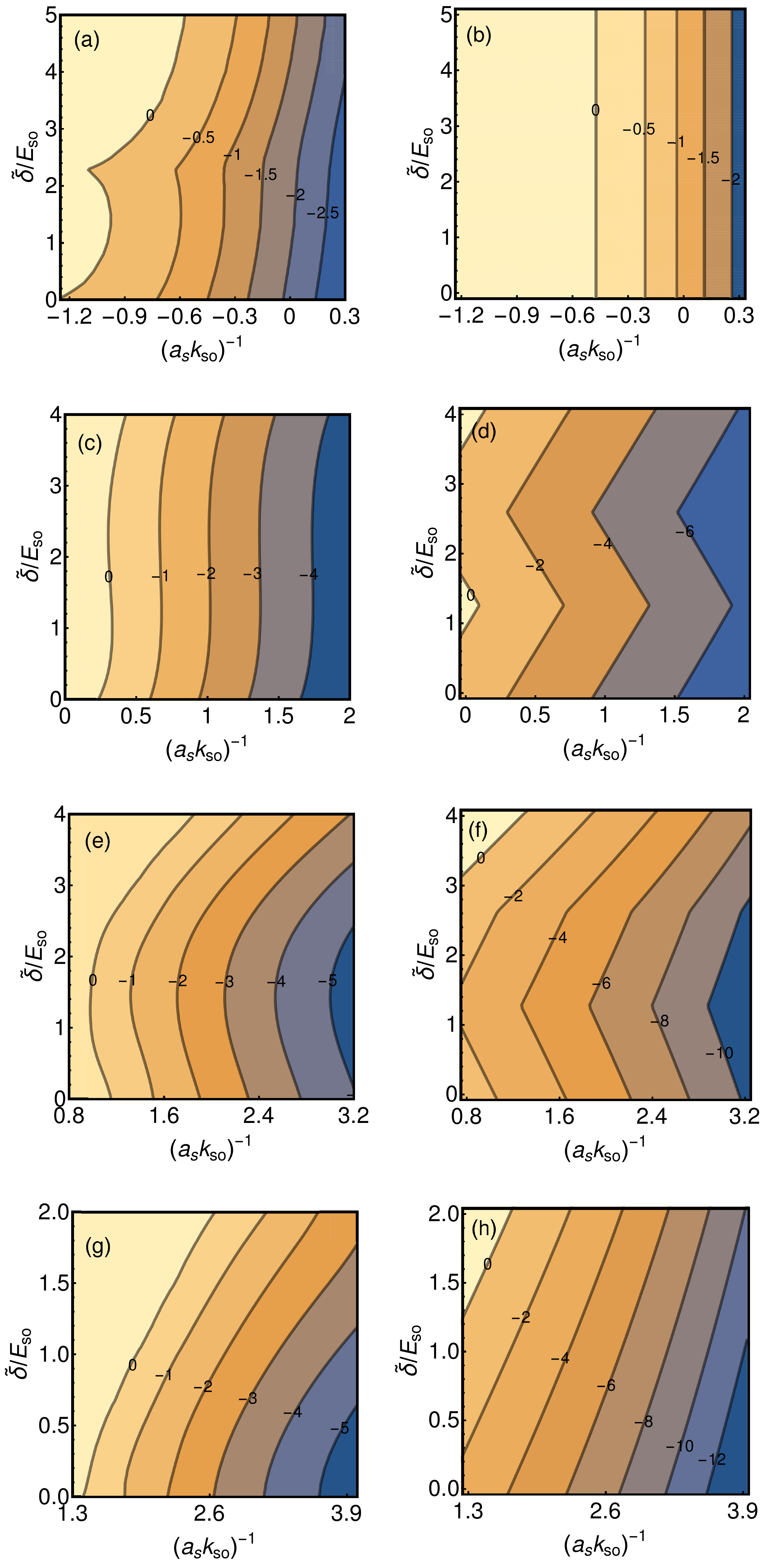}
\caption{(Color online)
The contours show the negative of the
three-boson binding energy, in units of $E_{\text{so}}$, 
as functions of $(a_s k_{\text{so}})^{-1}$ and $\tilde{\delta}/E_{\text{so}}$
for $k_{\text{so}}/\kappa_*=1.32$ and $\Omega=2E_{\text{so}}$ [panels~(a), (c), (e), and (g)] and infinitesimally small $\Omega$ [panels~(b), (d), (f), and (h)].
Panels in the first, second, third, and forth row 
show the negative of the binding energy for the three-boson states with $n=0, 1, 2$, and $3$, respectively.
Note the different ranges of the $x$-axis in panels~(a)-(h).
The four states $n=0-3$ correspond to the lowest Efimov manifold. 
Applying the generalized radial scaling law~\cite{GuanPRX2018},
these energy plots also describe higher-lying Efimov manifolds. 
}
\label{fig5}
\end{figure}

We start our discussion of the three-boson binding energy by considering the case for $\Omega=0$.
As for the two-boson system,
the $M_z$ quantum number of the three-boson system is conserved when $\Omega$ is equal to zero.
Since we have three bosons with pseudospin-$1/2$, 
the $M_z$ quantum number can take four different values, namely $-3/2, -1/2, 1/2$, and $3/2$.
Thus, for vanishing $\Omega$,
each energy curve in the ``normal'' Efimov scenario turns into an energy manifold 
consisting of four decoupled states.
If we measure the binding energies of these fixed $M_z$ states
with respect to the scattering threshold for the corresponding $M_z$ channel,
the binding energies of these states are the same as in the absence of spin-orbit coupling 
(see Appendix~\ref{appendix_omegazero}).
If we instead measure the binding energies for the $\Omega=0$ system with respect to the 
absolute minimum of all non-interacting relative dispersion curves, i.e.,
the lowest scattering threshold among the four different $M_z$ channels, 
the binding energies depend, in general, on $\tilde{\delta}$ (see Appendix~\ref{appendix_omegazero}).
As discussed in Sec.~\ref{sec_twobody}
in the context of the two-boson system, 
even though these binding energies are calculated with respect to the incorrect threshold, 
they can be interpreted as being those for infinitesimally small but finite $\Omega$.
The binding energies for infinitesimally small $\Omega$ are shown in Figs.~\ref{fig5}(b),~\ref{fig5}(d),~\ref{fig5}(f), and~\ref{fig5}(h)
for the lowest three-boson state $(n = 0)$, the first excited three-boson state $(n = 1)$, 
the second excited three-boson state $(n = 2)$, and the third excited three-boson state $(n=3)$, respectively.
The binding energy of the ground state is independent of the generalized detuning $\tilde{\delta}$ 
while the binding energies of the three excited states in the lowest energy manifold depend on $\tilde{\delta}$.

For finite $\Omega$,
the $M_z$ quantum number is not conserved any more. 
In this case,
the three-boson binding energies $\oline{E}_{3,\text{binding}}^{(n)}(\tilde{\delta})$ need to be determined numerically. 
Figures~\ref{fig5}(a),~\ref{fig5}(c),~\ref{fig5}(e), and~\ref{fig5}(g) 
show the negative of the binding energies $\oline{E}_{3,\text{binding}}^{(n)}(\tilde{\delta})$ 
for $n=0$, $n=1$, $n=2$, and $n=3$, respectively, for $\Omega=2E_{\text{so}}$ 
as functions of $(a_sk_{\text{so}})^{-1}$ and $\tilde{\delta}/E_{\text{so}}$.
Comparison of the binding energies for $\Omega=2E_{\text{so}}$ (left column of Fig.~\ref{fig5})
and those for infinitesimally small $\Omega$ (right column of Fig.~\ref{fig5}) shows that
the ``shapes'' of the contours in the same row are quite similar for $n=1-3$ but not for $n=0$.
For the same $(a_sk_{\text{so}})^{-1}$ and $\tilde{\delta}/E_{\text{so}}$,
the binding energies of the excited states for finite $\Omega$ are smaller than those for infinitesimally small $\Omega$.
In contrast,
the binding of the ground state is enhanced due to the presence of the spin-orbit coupling [compare Figs.~\ref{fig5}(a) and~\ref{fig5}(b)]. 
For example, 
the $\tilde{\delta}=0$ system with infinitesimally small $\Omega$ supports a bound state for $(a_sk_{\text{so}})^{-1}\ge -0.504$
while that with $\Omega=2E_{\text{so}}$ supports a bound state for $(a_sk_{\text{so}})^{-1}\ge -1.304$.
Figure~\ref{fig5}(a) shows that the binding energy for $\Omega=2E_{\text{so}}$ is enhanced the most 
for $\tilde{\delta}\approx 0$.
This implies that the critical generalized detuning $\tilde{\delta}_{\text{cr}}$
for $\Omega=2E_{\text{so}}$ is equal to zero.
In addition, 
Fig.~\ref{fig5}(a) displays a somewhat weaker enhancement for $\tilde{\delta}\approx 2.27E_{\text{so}}$.
For infinitesimally small $\Omega$ [see Fig.~\ref{fig5}(b)], in contrast, no such dependence on $\tilde{\delta}$ is observed; 
in this case, a bound state is supported for $(a_sk_{\text{so}})^{-1}\ge -0.504$ for all $\tilde{\delta}$.

\begin{figure}
\vspace*{0.cm}
\hspace*{0.cm}
\includegraphics[width=0.5\textwidth]{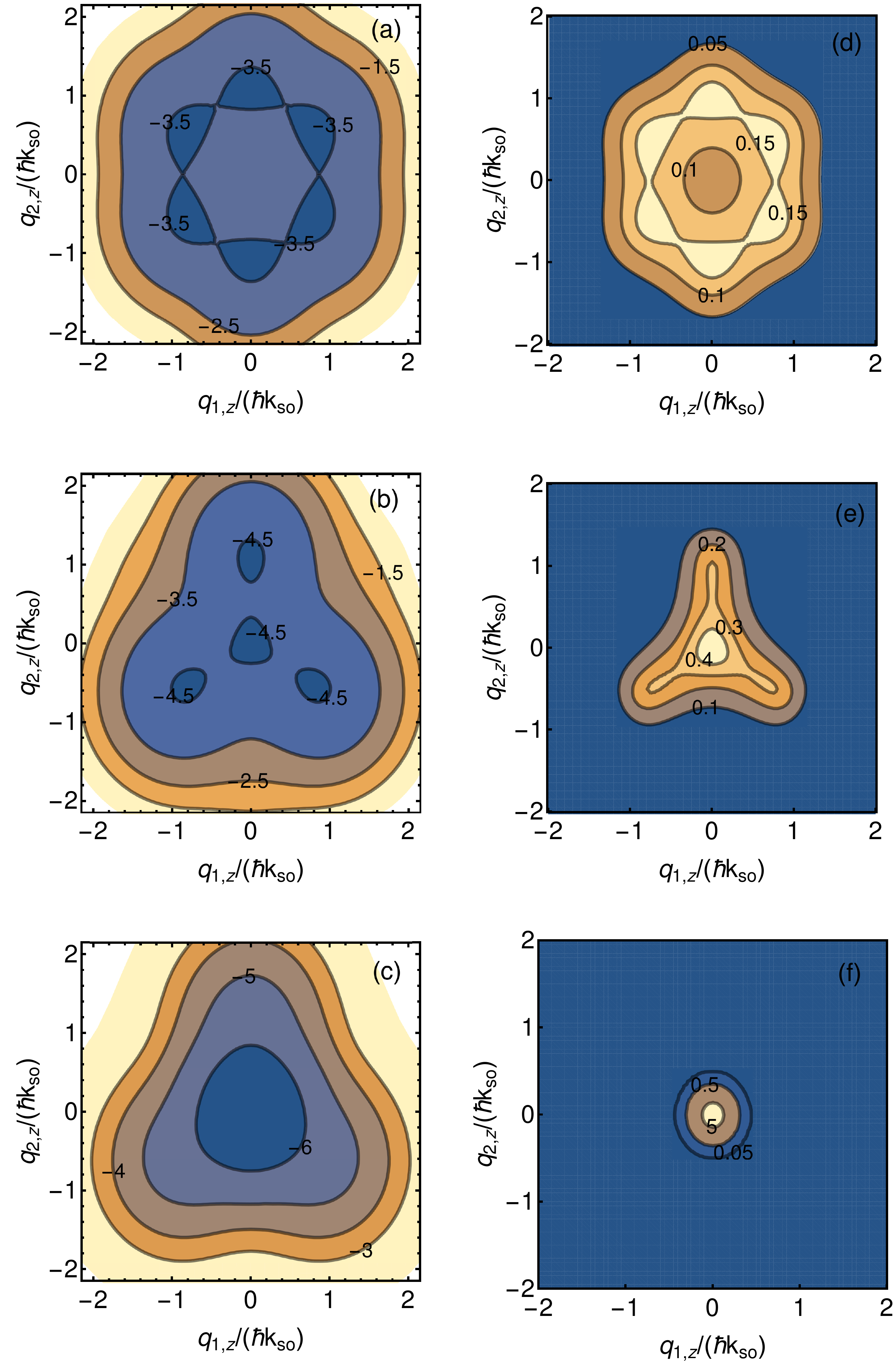}
\caption{(Color online)
Correlations between the lowest non-interacting relative dispersion curve and the relative momentum 
distribution for the ground state for $\Omega=2E_{\text{so}}$ and $(a_sk_{\text{so}})^{-1}=-0.96$.
Panels (a), (b), and (c) show the lowest non-interacting relative dispersion curves for $\tilde{\delta}=0$,
$2.27E_{\text{so}}$, and $3.5E_{\text{so}}$, respectively.
Panels (d), (e), and (f) show the relative momentum 
distributions for the ground state for $\tilde{\delta}=0$,
$2.27E_{\text{so}}$, and $3.5E_{\text{so}}$, respectively.
Note that the contours in panels (a)-(e) are equally spaced while
those in panel (f) are not. 
}
\label{fig6}
\end{figure}

To gain more insights into the weakly bound regime,
we look at the lowest non-interacting relative dispersion curve $\oline{E}^{(0)}_{3,\text{rel,ni}}(q_{1,z}, q_{2,z}, \tilde{\delta})$
and the relative momentum distribution $n(q_{1,z}, q_{2,z})$ 
of the three-boson ground state. 
The relative momentum distribution is defined through
\begin{eqnarray}
\label{three_body_momentum}
n(q_{1,z}, q_{2,z})=\sum_{\sigma}\int\Phi_{\text{rel},\sigma}^*(\vv{q}'_1, \vv{q}'_2)\delta(q_{1,z}'-q_{1,z})\times\\\nonumber
\delta(q_{2,z}'-q_{2,z})
\Phi_{\text{rel},\sigma}(\vv{q}'_1,\vv{q}'_2)d\vv{q}'_{1}d\vv{q}'_{2},
\end{eqnarray}
where $\sigma$ runs over the eight possible spin configurations. 
Figures~\ref{fig6}(a),~\ref{fig6}(b), and~\ref{fig6}(c) show $\oline{E}^{(0)}_{3,\text{rel,ni}}(q_{1,z}, q_{2,z},\tilde{\delta})$ 
as functions of $q_{1,z}$ and $q_{2,z}$ for $\Omega=2E_{\text{so}}$ and three different $\tilde{\delta}$.
For $\tilde{\delta}=0$ [Fig.~\ref{fig6}(a)],
$\oline{E}^{(0)}_{3,\text{rel,ni}}(q_{1,z}, q_{2,z},\tilde{\delta})$ has six degenerate global minima located away from $(q_{1,z}, q_{2,z})=(0,0)$. 
For finite $\tilde{\delta}$,
a local minimum appears at $(q_{1,z}, q_{2,z})=(0,0)$ and 
the degeneracy of the six minima located at $(q_{1,z}, q_{2,z})\ne (0,0)$ is broken. 
Three minima turn into global minima while the other three turn into local minima. 
For $\tilde{\delta}=2.27E_{\text{so}}$ [Fig.~\ref{fig6}(b)],
the energy associated with the local minimum located at $(q_{1,z}, q_{2,z})=(0,0)$ is degenerate with the energies
associated with the global minima located at $(q_{1,z}, q_{2,z})\ne (0,0)$. 
For this $\tilde{\delta}$,
the global minimum of $\oline{E}^{(0)}_{3,\text{rel,ni}}(q_{1,z}, q_{2,z}, \tilde{\delta})$ is four-fold degenerate. 
For larger $\tilde{\delta}$ [Fig.~\ref{fig6}(c)],
the global minimum of $\oline{E}^{(0)}_{3,\text{rel,ni}}(q_{1,z}, q_{2,z}, \tilde{\delta})$ is one-fold degenerate and located at $(q_{1,z}, q_{2,z})=(0,0)$.
In summary, as $\tilde{\delta}$ increases from 0 to $2.27E_{\text{so}}$ to larger values,
the number of global minima of $\oline{E}^{(0)}_{3,\text{rel,ni}}(q_{1,z}, q_{2,z},\tilde{\delta})$ 
changes from six to three to four to one. 
Correspondingly, 
the binding energy of the ground state takes on a global maximum for $\tilde{\delta}$ equal to 0
and a local maximum for $\tilde{\delta}$ approximately equal to $2.27E_{\text{so}}$.
As in the two-boson system, 
the enhancement of the binding of the ground state 
is correlated with the degeneracy of the global minimum 
of the non-interacting relative dispersion curves. 

Figures~\ref{fig6}(d)-\ref{fig6}(f) show $n(q_{1,z}, q_{2,z})$
for the same parameters as those used in Figs.~\ref{fig6}(a)-\ref{fig6}(c) and $(a_sk_{\text{so}})^{-1}=-0.96$.
For this scattering length, 
the three-boson threshold is given by the three-atom threshold for all $\tilde{\delta}$ values,
i.e., the corresponding two-boson system does not support a bound state.
Comparison between the left and the right columns of Fig.~\ref{fig6} shows that 
the number of peaks of $n(q_{1,z}, q_{2,z})$ is equal to the number of global minima of $\oline{E}^{(0)}_{3,\text{rel,ni}}(q_{1,z}, q_{2,z},\tilde{\delta})$.
The values of $q_{1,z}$ and $q_{2,z}$ for which $n(q_{1,z}, q_{2,z})$ is maximal
are, to a very good approximation, identical to those for which $\oline{E}^{(0)}_{3,\text{rel,ni}}(q_{1,z}, q_{2,z},\tilde{\delta})$ is minimal.

The spin and the momentum in spin-orbit coupled systems are locked. 
Using this together with the fact that the total wave function has to be symmetric under 
the exchange of any two bosons, 
we can, in a first order approximation, 
assign spin configurations to the global minima in Fig.~\ref{fig6}.
For a single-particle system with 1D spin-orbit coupling, 
a spin-up configuration prefers to have a negative momentum along the $z$-direction to lower the energy
while a spin-down configuration prefers to have a positive momentum to lower the energy.
Thus, 
a pair of parallel spins prefers to have vanishing relative momentum and finite center-of-mass momentum
while a pair of anti-parallel spins prefers to have a finite relative momentum and vanishing center-of-mass momentum.
Global minima in Fig.~\ref{fig6} that are shifted away from $(0,0)$ are associated with anti-parallel spin configurations.   
For $\tilde{\delta}=0$ [Fig.~\ref{fig6}(d)],
$n(q_{1,z}, q_{2,z})$ has six peaks located away from $(q_{1,z}, q_{2,z})=(0,0)$;
this indicates that 
the ground state is primarily a superposition of the six spin states that contain anti-parallel spin pairs,
namely $|\uparrow\uparrow\downarrow\rangle$,
$|\uparrow\downarrow\uparrow\rangle$, $|\downarrow\uparrow\uparrow\rangle$,
$|\uparrow\downarrow\downarrow\rangle$,
$|\downarrow\downarrow\uparrow\rangle$, and $|\downarrow\uparrow\downarrow\rangle$.
For $\tilde{\delta}>0$,
the ground state prefers to have more spin-down particles than spin-up particles.
For $\tilde{\delta}=2.27E_{\text{so}}$ [Fig.~\ref{fig6}(e)],
$n(q_{1,z}, q_{2,z})$ has four peaks, 
three are located at $(q_{1,z}, q_{2,z})\ne (0, 0)$ 
and one at $(q_{1,z}, q_{2,z})=(0,0)$;
this indicates that
the ground state is primarily a superposition 
of the spin configurations
$|\uparrow\downarrow\downarrow\rangle$,
$|\downarrow\downarrow\uparrow\rangle$, 
$|\downarrow\uparrow\downarrow\rangle$,
and $|\downarrow\downarrow\downarrow\rangle$.
For $\tilde{\delta}=3.5E_{\text{so}}$ [Fig.~\ref{fig6}(f)],
$n(q_{1,z}, q_{2,z})$ has one peak located at $(q_{1,z}, q_{2,z})=(0,0)$;
this indicates that 
the ground state primarily consists of three spin-down spins,
i.e., the dominant spin configuration is $|\downarrow\downarrow\downarrow\rangle$.

\subsection{Total ground state energy}

\begin{figure}
\vspace*{0.2cm}
\hspace*{0.cm}
\includegraphics[width=0.35\textwidth]{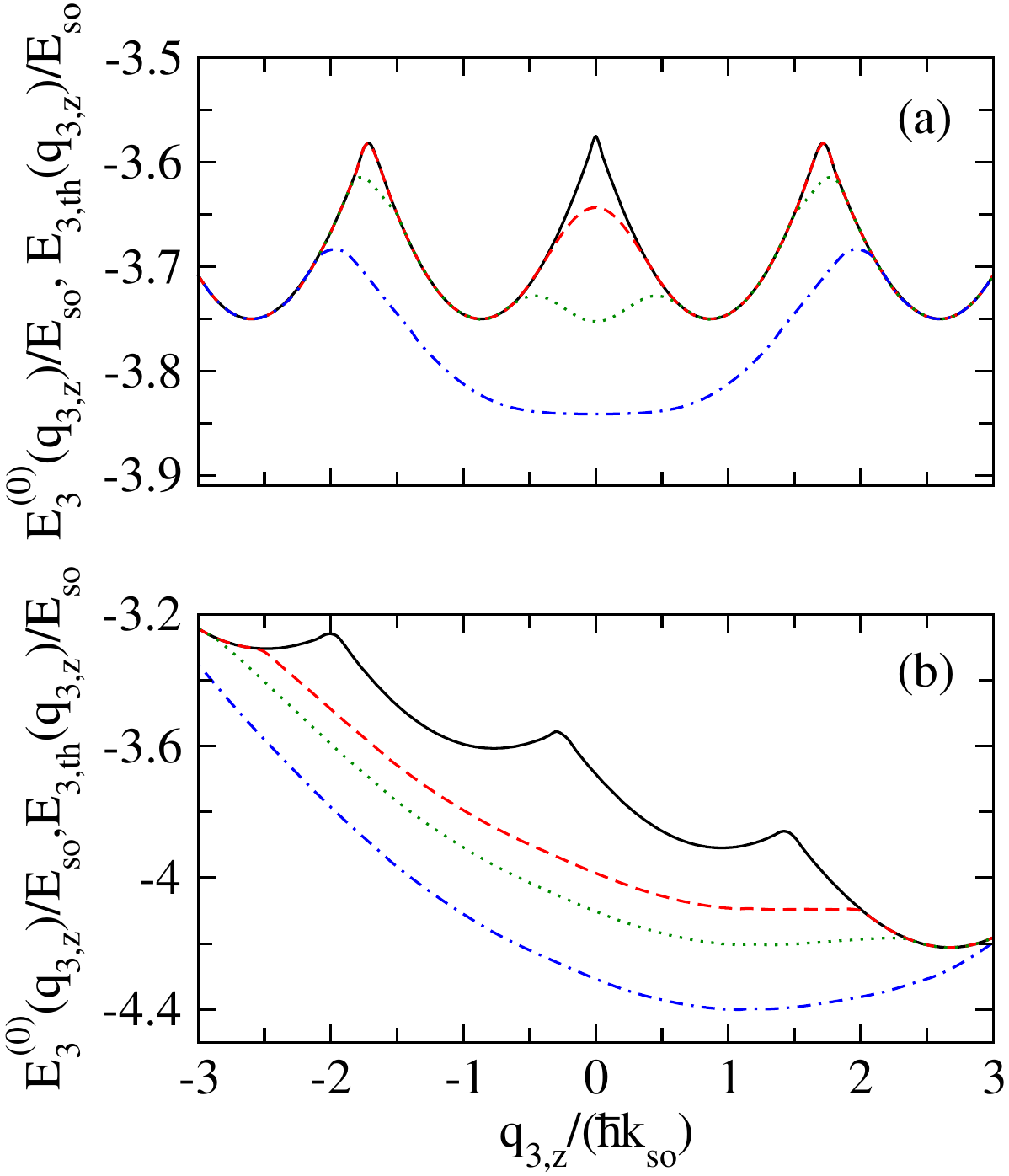}
\caption{(Color online)
The ground state energy $E_{3}^{(0)}(q_{3,z})$ of $\widehat{H}_3(q_{3,z})$ and the threshold energy $E_{3,\text{th}}(q_{3,z})$
for $\Omega=2E_{\text{so}}$ and various $(a_sk_{\text{so}})^{-1}$ and $\delta/E_{\text{so}}$ combinations. 
The black solid lines correspond to $E_{3,\text{th}}(q_{2,z})$.
The red dashed, green dotted, and blue dot-dashed lines in panel (a) show $E_{3}^{(0)}(q_{3,z})$
for $\delta=0$ and $(a_sk_{\text{so}})^{-1}=-1.1$, $-1.02$, and $-0.9$, respectively.
The red dashed, green dotted, and blue dot-dashed lines in panel (b) show $E_{3}^{(0)}(q_{3,z})$
for $\delta=0.35E_{\text{so}}$ and $(a_sk_{\text{so}})^{-1}=-0.8$, $-0.72$, and $-0.6$, respectively.
}
\label{fig7}
\end{figure}

This section discusses the total ground state energy $E_{3}^{(0)}(q_{3,z})$ of $\widehat{H}_{3}(q_{3,z})$,
which is obtained from $\oline{E}_{3,\text{rel}}^{(0)}(\tilde{\delta})$ using Eq.~\eqref{E_tot_2}.
Figure~\ref{fig7} shows examples for $\Omega=2E_{\text{so}}$ and 
various $(a_sk_{\text{so}})^{-1}$ and $\delta$ combinations.
For these parameter combinations, 
the three-boson threshold is given by the three-atom threshold
(i.e., two-boson bound states do not exist).
Correspondingly, 
the three-boson threshold, shown by black solid lines in Figs.~\ref{fig7}(a) and~\ref{fig7}(b)
for $\delta=0$ and $\delta=0.35E_{\text{so}}$, is independent of the value of $(a_sk_{\text{so}})^{-1}$. 
For vanishing $\delta$ [Fig.~\ref{fig7}(a)],
the scattering threshold $E_{3,\text{th}}(q_{3,z})$ has four global minima 
that are located at $q_{3,z}=\pm 0.866\hbar k_{\text{so}}$ and $q_{3,z}=\pm 2.598\hbar k_{\text{so}}$.
For $\delta=0.35E_{\text{so}}$ [Fig.~\ref{fig7}(b)],
the scattering threshold $E_{3,\text{th}}(q_{3,z})$ is ``tilted'' and
has one global minimum that is located at $q_{3,z}=2.671\hbar k_{\text{so}}$.
For both panels in Fig.~\ref{fig7},
the total ground state energy decreases (becomes more negative)
for decreasing $(a_sk_{\text{so}})^{-1}$.
Moreover,
for both panels the total ground state corresponds to a scattering state for the most negative $(a_sk_{\text{so}})^{-1}$
and to a bound state for $(a_sk_{\text{so}})^{-1}$ values larger than some critical value. 
As $(a_sk_{\text{so}})^{-1}$ changes for fixed $\delta/E_{\text{so}}$,
the degeneracy of the total ground state changes. 
For example, 
for vanishing $\delta$ [Fig.~\ref{fig7}(a)],
the total ground state is a four-fold degenerate scattering state located at finite $q_{3,z}$
for $(a_sk_{\text{so}})^{-1}=-1.1$ (red dashed line)
and a one-fold degenerate bound state located at vanishing $q_{3,z}$ for $(a_sk_{\text{so}})^{-1}=-0.9$ (blue dot-dashed line).
For $(a_sk_{\text{so}})^{-1}=-1.02$ (green dotted line),
the total ground state is five-fold degenerate:
four scattering states located at finite $q_{3,z}$ and one bound state located at vanishing $q_{3,z}$.
For $\delta=0.35E_{\text{so}}$ [Fig.~\ref{fig7}(b)],
the total ground state corresponds to a scattering state located at $q_{3,z}=2.671\hbar k_{\text{so}}$ for $(a_sk_{\text{so}})^{-1}=-0.8$ (red dashed line) and
a bound state located at $q_{3,z}\approx 1.187\hbar k_{\text{so}}$ for $(a_sk_{\text{so}})^{-1}=-0.6$ (blue dot-dashed line).
For $(a_sk_{\text{so}})^{-1}=-0.72$ (green dotted line),
the bound state located at $q_{3,z}\approx 1.166\hbar k_{\text{so}}$ is degenerate with the scattering state located at $q_{3,z}=2.671\hbar k_{\text{so}}$.

\begin{figure}
\vspace*{0.2cm}
\hspace*{0.cm}
\includegraphics[width=0.48\textwidth]{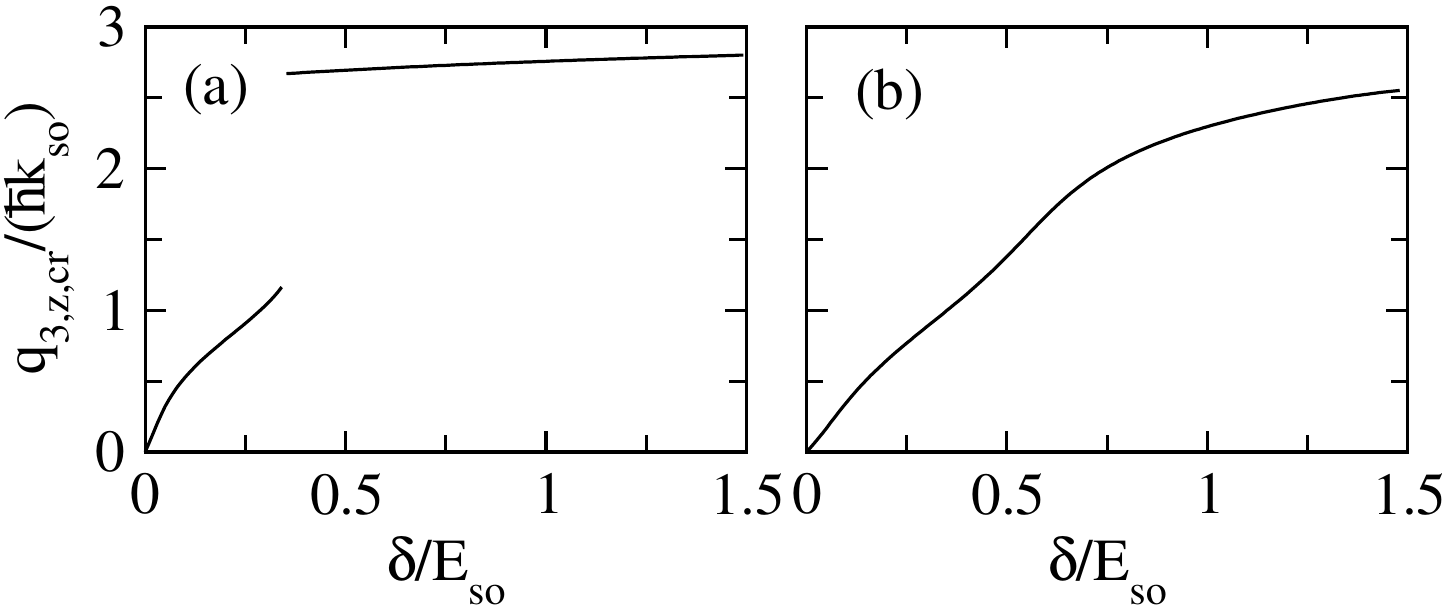}
\caption{
The critical center-of-mass momentum $q_{3,z,\text{cr}}$ 
of the three-boson ground state as a function of $\delta/E_{\text{so}}$
for $\Omega=2E_{\text{so}}$ and (a) $(a_sk_{\text{so}})^{-1}=-0.72$ and (b) $(a_sk_{\text{so}})^{-1}=-0.3$.
}
\label{fig8}
\end{figure}

Figures~\ref{fig8}(a) and~\ref{fig8}(b) show the critical generalized total momentum $q_{3,z,\text{cr}}$
as a function of $\delta/E_{\text{so}}$ for $(a_sk_{\text{so}})^{-1}=-0.72$ and $(a_sk_{\text{so}})^{-1}=-0.3$, respectively.
For the range of $\delta/E_{\text{so}}$ considered here ($0<\delta<1.5E_{\text{so}}$),
the three-boson threshold is equal to the three-atom threshold for $(a_sk_{\text{so}})^{-1}=-0.72$
and equal to the atom-dimer threshold for $(a_sk_{\text{so}})^{-1}=-0.3$.
For $(a_sk_{\text{so}})^{-1}=-0.72$ and $0<\delta<0.35E_{\text{so}}$,
$q_{3,z,\text{cr}}$ increases continuously from $0$ to $1.166\hbar k_{\text{so}}$.
In this regime, 
the total ground state corresponds to a bound state.
For $\delta=0.35E_{\text{so}}$,
the critical generalized total momentum $q_{3,z,\text{cr}}$ jumps from 
$q_{3,z}=1.166\hbar k_{\text{so}}$ to $q_{3,z}=2.671\hbar k_{\text{so}}$.
For $\delta>0.35E_{\text{so}}$, 
$q_{3,z,\text{cr}}$ increases very slowly. 
In this regime, 
the total ground state corresponds to a scattering state.
For $(a_sk_{\text{so}})^{-1}=-0.3$ [Fig.~\ref{fig8}(b)],
the total ground state corresponds to a bound state 
for all $\delta$ considered.
In this case, 
$q_{3,z,\text{cr}}$ increases continuously from $0$ to $2.56\hbar k_{\text{so}}$
as $\delta$ increases from $0$ to $1.49E_{\text{so}}$.
Thus, 
for the parameter combinations considered in this work,
the critical generalized total momentum $q_{3,z,\text{cr}}$ of the three-boson system 
varies continuously with respect to $\delta$ 
if the total ground state corresponds to a bound state 
and changes discontinuously 
if the total ground state jumps from a bound state to a scattering state. 

\begin{figure}
\vspace*{0.2cm}
\hspace*{0.cm}
\includegraphics[width=0.40\textwidth]{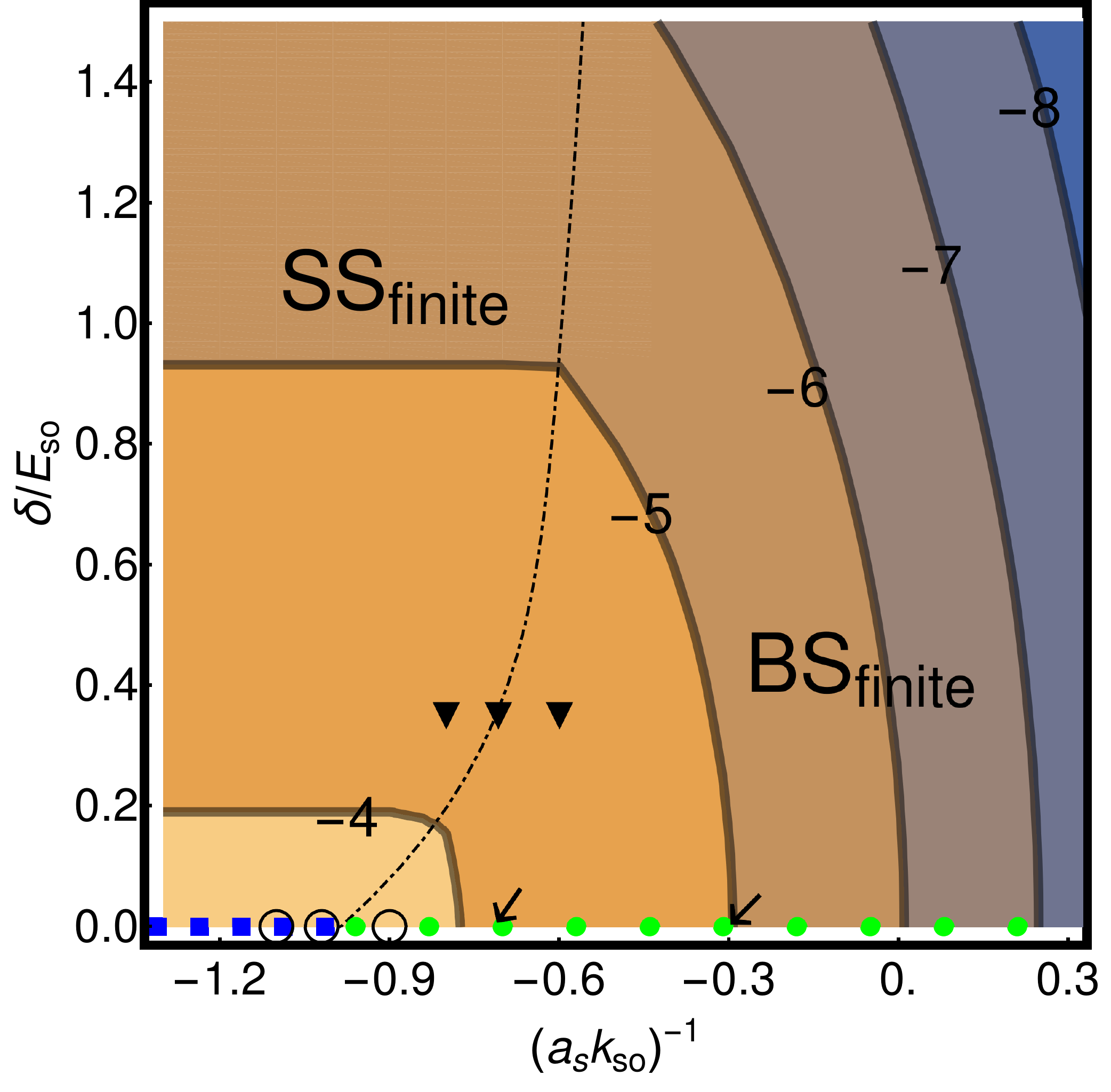}
\caption{(Color online)
``Phase diagram'' for the total ground state
as functions of $(a_sk_{\text{so}})^{-1}$ and $\delta/E_{\text{so}}$ for $\Omega=2E_{\text{so}}$.
The contours show the total ground state energy $E_{3}^{(0)}(q_{3,z})/E_{\text{so}}$.
The black dot-dashed line, the green circles, and the blue squares separate the two phases $\text{SS}_{\text{finite}}$
and $\text{BS}_{\text{finite}}$ from each other. 
The open circles and triangles mark the parameter combinations corresponding to Figs.~\ref{fig7}(a) and~\ref{fig7}(b).
The arrows mark the scattering lengths corresponding to Figs.~\ref{fig8}(a) and~\ref{fig8}(b).
}
\label{fig9}
\end{figure}

We repeat the analysis illustrated in Figs.~\ref{fig7} and~\ref{fig8} 
for other $(a_sk_{\text{so}})^{-1}$ and $\delta/E_{\text{so}}$ combinations
and summarize the results in the ``phase diagram'' for the total ground state
in Fig.~\ref{fig9} as functions of $(a_sk_{\text{so}})^{-1}$ and $\delta/E_{\text{so}}$ for $\Omega=2E_{\text{so}}$.
For this $\Omega/E_{\text{so}}$,
the total ground state falls in one of the following two phases.
$\text{SS}_{\text{finite}}$: The total ground state is a scattering state with $q_{3,z, \text{cr}}\ne 0$.
$\text{BS}_{\text{finite}}$: The total ground state is a bound state with $q_{3,z,\text{cr}}\ne 0$.
The region encircled by the blue squares, black dot-dashed line, and the upper and left edge of the figure 
corresponds to the phase $\text{SS}_{\text{finite}}$. 
The region encircled by the green circles, the right and upper edge of the figure, and the black dot-dashed line
corresponds to the phase $\text{BS}_{\text{finite}}$. 
Along the black dot-dashed line
[the green dotted line in Fig.~\ref{fig7}(b) corresponds to such a situation], 
the total ground state is two-fold degenerate:
a bound state located at finite $q_{3, z, \text{cr}}$ 
and a scattering state located at finite $q_{3, z, \text{cr}}$.
Along the green circles
[the blue dot-dashed line in Fig.~\ref{fig7}(a) corresponds to such a situation], 
the total ground state is a one-fold degenerate bound state with vanishing $q_{3, z, \text{cr}}$.
Along the blue squares,
the total ground state corresponds to a four-fold degenerate scattering state located at finite $q_{3, z, \text{cr}}$. 
In the phase $\text{BS}_{\text{finite}}$,
$q_{3,z,\text{cr}}$ of the total ground state 
changes smoothly with $\delta/E_{\text{so}}$
[see Fig.~\ref{fig8}(b)].
When the system crosses the black dot-dashed line,
$q_{3,z,\text{cr}}$ changes discontinuously
[see Fig.~\ref{fig8}(a)]. 
As pointed out in the context of studies of fermionic systems~\cite{ZhaiPRL2011, ZhaiRev2012, PuPRA2013, PuPRA2014, ShenoyPRA2013, JiangPRA2012, HuPRA2013, HuPRA2013_2},
the phase diagrams shown in Figs.~\ref{fig4} and~\ref{fig9} can provide useful input
to understand pairing mechanisms and dynamical properties of 
few-body clusters embedded in a Bose gas at zero temperature in the presence of 1D spin-orbit coupling.

\section{Conclusions}
\label{sec_conclusion}

The binding energy $\oline{E}^{(n)}_{3,\text{binding}}$ of the three-boson system 
in the presence of 1D spin-orbit coupling 
and short-range two-body $s$-wave interactions
depends on five independent parameters, namely, 
the spin-orbit coupling strength $k_{\text{so}}$, 
the Raman coupling $\Omega$, 
the generalized detuning $\tilde{\delta}$, 
the two-body $s$-wave scattering length $a_s$, 
and the three-body parameter $\kappa_*$.
Using the continuous scale invariance of the system~\cite{GuanPRX2018, shi2014, shi2015},
one parameter can be chosen as unit
and thereby ``scaled away''. 
In this work,
$k_{\text{so}}^{-1}$ and $E_{\text{so}}$ were chosen as length and energy units 
and all the results were represented in terms of dimensionless parameters.
The dimensionless binding energy $\oline{E}^{(n)}_{3,\text{binding}}/E_{\text{so}}$ 
depends on four independent dimensionless parameters, 
namely, 
$\Omega/E_{\text{so}}$, $\tilde{\delta}/E_{\text{so}}$, 
$a_sk_{\text{so}}$, and $\kappa_*/k_{\text{so}}$.
According to the generalized radial scaling law~\cite{GuanPRX2018},
once the full parameter-dependence of the dimensionless binding energy surfaces
$\oline{E}^{(n)}_{3,\text{binding}}/E_{\text{so}}$ 
are mapped out for one energy manifold, 
one can radially rescale the binding energy surfaces in
the $(\oline{E}^{(n)}_{3,\text{binding}}/E_{\text{so}}, \Omega/E_{\text{so}}, \tilde{\delta}/E_{\text{so}}, a_sk_{\text{so}}, \kappa_*/k_{\text{so}})$ space
by discrete scaling factors to obtain the entire spectrum.

This work mapped out the three-boson binding energy surfaces
for the lowest universal energy manifold 
in a subspace that is characterized by $\Omega=2E_{\text{so}}$ and $\kappa_*\approx 0.758k_{\text{so}}$.
For the parameter combinations considered,
the two- and three-boson systems share some trends despite the fact that the three-boson system depends on 
$\kappa_*/k_{\text{so}}$ while the two-boson system does not.  
For example, 
the shapes of the energy surfaces for $\Omega=2E_{\text{so}}$ are similar to those for infinitesimally small $\Omega$.
For the two-boson system,
the binding energy of the ground state is enhanced 
while the binding energies of the first and second excited states are weakened
in the presence of 1D spin-orbit coupling.
Similarly,
for the three-boson system,
the binding energy of the ground state in the lowest energy manifold is enhanced 
while the binding energies of the first, second, and third excited states are weakened
in the presence of 1D spin-orbit coupling.
The enhancement of the binding energies of the ground state of the two- and three-boson systems
is correlated with the degeneracy of the lowest scattering threshold;
this is consistent with the density-of-state argument presented in Ref.~\cite{ShenoyPRA2013} for two spin-orbit coupled fermions.
The modification of the binding energies due to the spin-orbit coupling is associated with a modification of the 
critical scattering lengths at which dimers and trimers merge with the lowest scattering threshold. 
Taking advantage of Feshbach resonance tuning~\cite{EiteRMP2010}, 
the critical scattering lengths can be measured 
by monitoring atom losses in cold atom experiments~\cite{grimm2006, Zaccanti, ferlaino2010, ferlaino2011, grimm2011, huangPRL}. 
The enhanced spin-orbit coupling induced binding is also expected to play a role in the context of many-body physics. 
For example, 
a flow-enhanced pairing induced by the 1D spin-orbit coupling in a Fermi gas is discussed in Ref.~\cite{ShenoyPRA2013}.

In the weakly bound regime, 
the relative momentum distributions of the dimer and trimer show rich structures in the presence of 1D spin-orbit coupling. 
These structures are correlated with the structures of the corresponding lowest relative dispersion curves. 
Due to spin-momentum locking, 
the weakly bound states have various mixtures of different spin configurations. 
The characteristics of the relative momentum distribution and spin structure are expected to be measurable 
in dedicated cold atom experiments. 

Due to the breaking of the Galilean invariance in 1D spin-orbit coupled systems,
the eigenenergies of the full Hamiltonian 
depend non-trivially on the generalized total momentum. 
This is in contrast to the corresponding systems without spin-orbit coupling.
We determined the ``phase diagram'' of the total ground state of two- and three-boson systems.
In these phase diagrams, 
the phase boundaries separate phases that are characterized by different generalized total momenta.
These phase diagrams provide guidance for many-body studies. 
The generalized total momentum of 1D spin-orbit coupled systems,
realized using the Raman laser scheme in cold atom systems~\cite{SpielmanNature2011, galitski_review2013, goldman2014},
can be measured 
using time-of-flight imaging.
In contrast to the generalized total momentum, 
the total mechanical momentum of the total ground state vanishes
regardless of the values of the spin-orbit coupling parameters. 

In the future, 
it will be interesting to explore the analogous physics 
for systems consisting of three fermions, 
away and in the vicinity of a two-body $p$-wave resonance.     
Since the dimer and trimer energies depend on the generalized total momentum,
it would also be interesting to study 
three-body systems in the presence of spin-orbit coupling and an external harmonic trap.
For these systems, 
states with different total generalized momentum are coupled.

\section{Acknowledgement}
\label{acknowledgement}
Support by the National Science Foundation through
grant number  
PHY-1806259
is gratefully acknowledged.
This work used the Extreme Science and Engineering
Discovery Environment (XSEDE), which is supported by
NSF Grant No. OCI-1053575,
and the OU
Supercomputing Center for Education and Research 
(OSCER) at the University of Oklahoma (OU).

\appendix

\section{Analytical solutions for $\Omega=0$}
\label{appendix_omegazero}
This section provides analytical solutions for 
interacting systems with
$\Omega=0$, $\tilde{\delta} \ge 0$, and $k_{\text{so}} \ne 0$
(throughout we assume $\tilde{\delta} \ge 0$; the negative $\tilde{\delta}$
case can be treated analogously).
For $\Omega=0$, 
there exists no
coupling between the different spin states and the relative
dispersion curves of the non-interacting system
and the eigenstates of the interacting system
can be labeled by the
product spin states, i.e., by the $M_z$ projection quantum number.
The eigenstates of the interacting
system with $\Omega=0$, $\tilde{\delta} \ge 0$,
and $k_{\text{so}}\ne0$ can be determined analytically
provided the solutions of the corresponding 
interacting system for $\Omega=\delta=k_{\text{so}}=0$ are known.
Although the $\Omega=0$ Hamiltonian with $k_{\text{so}} \ne 0$ is an artificial construct, 
its solutions provide a good deal of guidance for the $\Omega \ne 0$ Hamiltonian.

We start with the two-boson system interacting through
a zero-range two-body potential with 
positive $s$-wave scattering length $a_s$.
Denoting the relative wave function of 
the system with $\Omega=\delta=k_{\text{so}}=0$
by $\psi_{2,\text{sr}}(|\vv{\rho}_{1}|)$ and the corresponding 
relative eigenenergy
by $E_{2,\text{sr}}$ (``sr'' stands for ``short-range''),
three symmetric eigenstates of the two-boson system with $k_{\text{so}} \ne 0$ and $\tilde{\delta} \ge 0$
can be constructed,
\begin{eqnarray}
\Psi_{2, -1}=\psi_{2,\text{sr}}(|\vv{\rho}_{1}|) 
\exp \left( \frac{\imath}{\hbar} \vv{q}_2 \cdot \vv{\rho}_2 \right)
| \downarrow \downarrow \rangle,
\end{eqnarray}
\begin{eqnarray}
\Psi_{2, 1}=\psi_{2,\text{sr}}(|\vv{\rho}_{1}|) 
\exp \left( \frac{\imath}{\hbar} \vv{q}_2 \cdot \vv{\rho}_2 \right)
| \uparrow \uparrow \rangle,
\end{eqnarray}
and
\begin{eqnarray}
\Psi_{2, 0}=
\frac{\psi_{2,\text{sr}}(|\vv{\rho}_{1}|)}{\sqrt{2}} 
\exp \left( \frac{\imath}{\hbar} \vv{q}_2 \cdot \vv{\rho}_2 \right)
\times \nonumber \\
\left[
\exp( -\imath k_{\text{so}} {\rho}_{1,z}) | \uparrow \downarrow \rangle
+
\exp( \imath k_{\text{so}} \rho_{1,z}) | \downarrow \uparrow \rangle
\right].
\end{eqnarray}
Here, the first subscript of $\Psi$ denotes the particle number and the second subscript the $M_z$
quantum number. 
The corresponding eigenenergies of $\widehat{H}_2$ are
\begin{eqnarray}
E_{2,-1}=E_{2, \text{sr}}-\tilde{\delta} + \frac{\vv{q}_2^2}{2 \mu_2},
\end{eqnarray}
  \begin{eqnarray}
E_{2, 1}=E_{2, \text{sr}}+\tilde{\delta}+ \frac{\vv{q}_2^2}{2 \mu_2},
\end{eqnarray}
and
  \begin{eqnarray}
E_{2, 0}=E_{2, \text{sr}}-2 E_{\text{so}}+ \frac{\vv{q}_2^2}{2 \mu_2},
\end{eqnarray}
respectively.

For $\Omega=0$, the binding energy is obtained by
referencing the eigenenergy relative to the
atom-atom threshold energy that is associated
with a state that has the same
$M_z$ projection quantum number as the state considered.
Doing so yields a binding energy of
$|E_{2,\text{sr}}|$ for all three two-boson states;
these bound states exist provided $\psi_{2,\text{sr}}$ describes a bound state.

Next, we consider the three-particle system.
For $\Omega=\tilde{\delta}=k_{\text{so}}=0$, the 
energies of three identical bosons
with zero- or short-range two-body interactions
have been---building on the seminal work by
Efimov~\cite{efimov70}---studied extensively.
Denoting the relative three-boson eigen state 
for $\Omega=\tilde{\delta}=k_{\text{so}}=0$ 
by $\psi_{3,\text{sr}}(\vv{\rho}_1,\vv{\rho}_2)$ and the corresponding
eigen energy by $E_{3,\text{sr}}$
(the state considered can be any one of the Efimov states for,
at this point, unspecified scattering length),
four fully symmetric eigen states of the Hamiltonian $\widehat{H}_3$ with finite
$k_{\text{so}}$ and vanishing $\Omega$ can be constructed,
\begin{eqnarray}
\label{eq_state3_nosoc1}
\Psi_{3,3/2}=\psi_{3,\text{sr}}(\vv{\rho}_1,\vv{\rho}_2) 
\exp \left( \frac{\imath}{\hbar} \vv{q}_3 \cdot \vv{\rho}_3 \right)
| \uparrow \uparrow \uparrow \rangle,
\end{eqnarray}
\begin{eqnarray}
\label{eq_state3_nosoc2}
\Psi_{3,-3/2}=\psi_{3,\text{sr}} (\vv{\rho}_1,\vv{\rho}_2) 
\exp \left( \frac{\imath}{\hbar} \vv{q}_3 \cdot \vv{\rho}_3 \right)
| \downarrow \downarrow \downarrow \rangle,
\end{eqnarray}
\begin{eqnarray}
\label{eq_state3_nosoc3}
\Psi_{3,1/2}=&&
\frac{\psi_{3,\text{sr}}(\vv{\rho}_1,\vv{\rho}_2) }{\sqrt{3}}
\exp \left( \frac{\imath}{\hbar} \vv{q}_3 \cdot \vv{\rho}_3 \right) \times \nonumber \\
\Big[ 
&& \exp \left( - \frac{ \imath 4}{3} k_{\text{so}} z_{12,3} \right) | \uparrow \uparrow \downarrow \rangle
+ \nonumber \\
&& \exp \left( - \frac{ \imath 4}{3} k_{\text{so}} z_{13,2} \right) | \uparrow \downarrow \uparrow \rangle
+ \nonumber \\
&& \exp \left( - \frac{ \imath 4}{3} k_{\text{so}} z_{23,1} \right) | \downarrow \uparrow \uparrow \rangle
\Big],
\end{eqnarray}
and
\begin{eqnarray}
\label{eq_state3_nosoc4}
\Psi_{3,-1/2}=&&
\frac{\psi_{3,\text{sr}}(\vv{\rho}_1,\vv{\rho}_2) }{\sqrt{3}}
\exp \left( \frac{\imath}{\hbar} \vv{q}_3 \cdot \vv{\rho}_3 \right) \times \nonumber \\
\Big[ 
&& \exp \left( \frac{ \imath 4}{3} k_{\text{so}} z_{12,3} \right) | \downarrow \downarrow \uparrow \rangle
+ \nonumber \\
&& \exp \left( \frac{ \imath 4}{3} k_{\text{so}} z_{13,2} \right) | \downarrow \uparrow \downarrow \rangle
+ \nonumber \\
&& \exp \left( \frac{ \imath 4}{3} k_{\text{so}} z_{23,1} \right) | \uparrow \downarrow \downarrow \rangle
\Big].
\end{eqnarray}
Here, $z_{ij,k}$ is defined as $(r_{i,z}+r_{j,z})/2-r_{k,z}$,
with $r_{i,z}$ denoting the $z$-component of the $i$th position
vector $\vv{r}_{i}$.
The corresponding eigenenergies are 
\begin{eqnarray}
\label{eq_en3_nosoc1}
E_{3, 3/2}=E_{3,\text{sr}}+ \frac{3 \tilde{\delta}}{2} + \frac{\vv{q}_3^2}{2 \mu_3},
\end{eqnarray}

\begin{eqnarray}
\label{eq_en3_nosoc2}
E_{3, -3/2}=E_{3,\text{sr}}- \frac{3 \tilde{\delta}}{2} + \frac{\vv{q}_3^2}{2 \mu_3},
\end{eqnarray}
\begin{eqnarray}
\label{eq_en3_nosoc3}
E_{3,1/2}=E_{3,\text{sr}}- \frac{8 E_{\text{so}}}{3} +  \frac{\tilde{\delta}}{2} + 
\frac{\vv{q}_3^2}{2 \mu_3},
\end{eqnarray}
and
\begin{eqnarray}
\label{eq_en3_nosoc4}
E_{3,-1/2}=E_{3, \text{sr}}- \frac{8 E_{\text{so}}}{3} -  \frac{\tilde{\delta}}{2} + 
\frac{\vv{q}_3^2}{2 \mu_3}.
\end{eqnarray}
Equations~(\ref{eq_state3_nosoc1})-(\ref{eq_en3_nosoc4})
apply to every three-boson state,
i.e., for each 
$s$-wave scattering length a given three-boson state
 $\psi_{3,\text{sr}}$
is ``split'' into four states.
As in the two-boson case,
the binding energy for $\Omega=0$
is obtained by
referencing the eigen energy relative to the
three-boson threshold energy that is associated with a state
that has the same
$M_z$ projection quantum number as the state considered.
Assuming two-body zero-range $s$-wave interactions,
doing so yields a binding energy of
$|E_{3,\text{sr}}|$ for $a_s<0$ 
and a binding energy of $|E_{3,\text{sr}}-E_{2,\text{sr}}|$
for $a_s>0$ for all four three-boson states
(these bound states exist provided the state $\psi_{3,\text{sr}}$
describes a bound state).

An infinitesimally small $\Omega$ introduces couplings between 
the different product spin states.
As a consequence, $M_z$
is no longer a good quantum number
and the interacting states for
infinitesimally small $\Omega$ have a finite overlap with 
the state(s) that is (are)
associated with the lowest two- or three-boson thresholds
(lowest energy of {\em{all}} the non-interacting relative dispersion curves).
To get a first sense of how this impacts the binding energies, 
Tables~\ref{tab_2body} and \ref{tab_trivial3} summarize the binding energies
for infinitesimally
small $\Omega$, 
calculated using the $\Omega =0$ energies
reported above and referencing the energies
relative to the lowest two- and three-boson threshold, respectively.
These binding energies are shown in Figs.~\ref{fig_ebind_twobody} and
\ref{fig5} of the main text.

\begin{widetext}

\begin{table}
  \begin{center}
  \begin{tabular}{l|c|c}
    & $0 \le \tilde{\delta} \le 2 E_{\text{so}}$ & $2 E_{\text{so}} < \tilde{\delta}$ \\
    \hline
    $M_z \approx 1$ & $|E_{2,\text{sr}}|- 2 E_{\text{so}} - \tilde{\delta}$ for $(a_s)^{-1} \ge \sqrt{2 \mu_1 (2 E_{\text{so}} + \tilde{\delta})} \hbar$ & $|E_{2,\text{sr}}|-2E_{\text{so}}-\tilde{\delta}$ for $(a_s)^{-1} \ge \sqrt{2 \mu_1 (2 E_{\text{so}} + \tilde{\delta})} / \hbar$ \\
    $M_z \approx -1$ & $|E_{2,\text{sr}}|- 2 E_{\text{so}} + \tilde{\delta}$ for $(a_s)^{-1} \ge  \sqrt{2 \mu_1 (2 E_{\text{so}} - \tilde{\delta})} / \hbar$ & $|E_{2,\text{sr}}|$ for $(a_s)^{-1} \ge 0$  \\
    $M_z \approx 0$ & $|E_{2,\text{sr}}|$ for $(a_s)^{-1} \ge 0$ & $|E_{2,\text{sr}}|+2E_{\text{so}} - \tilde{\delta}$ for $(a_s)^{-1} \ge \sqrt{2 \mu_1 (\tilde{\delta}- 2 E_{\text{so}})} / \hbar$  \\
  \end{tabular}
  \caption{Two-boson binding energies 
    for infinitesimally small
    Raman coupling strength $\Omega$
and  generalized detuning $\tilde{\delta}$ greater or equal to zero.
Column 1 lists the two-boson state considered
(the $M_z$ quantum numbers are approximate
since $\Omega$ is assumed to be finite).
Columns 2 to 3 list the corresponding
binding energies. It is assumed that the two-boson
system interacts through a zero-range potential.
For $\Omega=\delta=k_{\text{so}}=0$,
a single bound state with
energy $E_{2,\text{sr}}$ [binding energy
$|E_{2,\text{sr}}|=\hbar^2/(2 \mu_1 a_s^2)$] is supported for positive $a_s$.} 
\label{tab_2body}
  \end{center}
  \end{table}

\begin{table}
  \begin{center}
  \begin{tabular}{l|c|c}
    & $0 \le \tilde{\delta}/E_{\text{so}} \le 8/3, a_s \le 0$  
& $8/3 < \tilde{\delta}/E_{\text{so}}, a_s \le 0$  \\
   \hline
   $M_z \approx 3/2$ & $|E_{3,\text{sr}}| -  8 E_{\text{so}}/3- 2 \tilde{\delta} $ for $|E_{3,\text{sr}}| \ge 2 \tilde{\delta} + 8 E_{\text{so}}/3$ & $|E_{3,\text{sr}}| - 3 \tilde{\delta}$
   for $E_{3,\text{sr}}| \ge 3 \tilde{\delta}$ \\
$M_z \approx -3/2$ & $|E_{3,\text{sr}}| - 8 E_{\text{so}}/3+ \tilde{\delta} $ for $|E_{3,\text{sr}}| \ge 8 E_{\text{so}}/3 - \tilde{\delta}$ & $|E_{3,\text{sr}}|$ \\
$M_z \approx 1/2$ & $|E_{3,\text{sr}}| - \tilde{\delta}$  for $|E_{3,\text{sr}}| \ge \tilde{\delta}$ & $|E_{3,\text{sr}}|+  8 E_{\text{so}}/3 - 2 \tilde{\delta} $ for $|E_{3,\text{sr}}| \ge 2 \tilde{\delta} - 8 E_{\text{so}}/3$\\
$M_z \approx -1/2$ & $|E_{3,\text{sr}}|$ & $|E_{3,\text{sr}}|+  8 E_{\text{so}}/3 -  \tilde{\delta} $ for $|E_{3,\text{sr}}| \ge \tilde{\delta} - 8 E_{\text{so}}/3$ 
  \end{tabular}
  \caption{Three-boson binding energies for infinitesimally small Raman coupling strength $\Omega$
and generalized detuning $\tilde{\delta}$ greater or equal to zero.
Column 1 lists the three-boson state considered
(the $M_z$ quantum numbers are approximate
since $\Omega$ is assumed to be finite). Columns 2 and 3 list the corresponding
binding energies (the entries apply to any energy of the three-boson
Efimov plot) assuming 
two-body zero-range
interactions with negative $a_s$.
The energy of the
three-boson system with zero-range interactions
for $\Omega=\delta=k_{\text{so}}=0$ is denoted by
$E_{3,\text{sr}}$.
For $\Omega=\delta=k_{\text{so}}=0$,
the two-boson system supports a
single bound state
with energy $E_{2,\text{sr}}$ for positive $a_s$ but not for negative $a_s$.
To obtain the three-boson binding energies for positive $a_s$,
the quantity $-|E_{2,\text{sr}}|$ has to be added to the entries given
in columns 2 and 3.
}
\label{tab_trivial3}
  \end{center}
  \end{table}

\end{widetext}


\begin{thebibliography}{100}

\bibitem{efimov70}
V. Efimov,
Energy levels
  arising from resonant two-body forces in a three-body system,
Phys. Lett. B {\bf{33}}, 563 (1970).

\bibitem{efimov71}
V. Efimov,
Weakly-bound states of three resonantly-interacting particles, Yad. Fiz. {\bf{12}}, 1080-91 (1970).

\bibitem{efimov72}
V. Efimov, 
Level spectrum of three resonantly interacting particles, 
Sov. Phys. JETP Lett. {\bf{16}}, 34 (1972).

\bibitem{efimov73}
V. Efimov, 
Energy levels of three resonantly interacting particles, 
Nucl. Phys. A {\bf{210}}, 157 (1973).


\bibitem{braaten2006}
E. Braaten and H.-W. Hammer,
Universality
  in few-body systems with large scattering length,
Phys. Rep. {\bf{428}}, 259 (2006).

\bibitem{grimm2006}
T. Kraemer, M. Mark, P. Waldburger, J. G. Danzl, C. Chin, B. Engeser, 
A. D. Lange, K. Pilch, A. Jaakkola, H.-C. N\"agerl, and R. Grimm,
Evidence for Efimov quantum states in an ultracold gas of caesium atoms,
Nature (London) {\bf{440}}, 315 (2006).

\bibitem{Zaccanti}
M. Zaccanti, B. Deissler, C. D'Errico, M. Fattori, M. Jona-Lasinio, S. M\"uller, G. Roati, M. Inguscio, G. Modugno,
Observation of an Efimov spectrum in an atomic system,
Nat. Phys. {\bf{5}}, 586 (2009).

\bibitem{ferlaino2010}
F. Ferlaino and R. Grimm,
Trend: Forty years of Efimov physics: 
How a bizarre prediction turned into a hot topic,
Physics {\bf{3}}, 9 (2010).

\bibitem{ferlaino2011}
F. Ferlaino, A. Zenesini, M. Berninger, B. Huang, H.-C. N\"agerl,
and R. Grimm,
Efimov Resonances in Ultracold Quantum Gases,
Few-Body Syst. {\bf{51}}, 113 (2011).

\bibitem{grimm2011}
M. Berninger, A. Zenesini, B. Huang, W. Harm, H.-C. N\"agerl, F. Ferlaino, R. Grimm, P. S. Julienne, and J. M. Hutson,
Universality of the Three-Body Parameter for Efimov States in Ultracold Cesium,
Phys. Rev. Lett. {\bf{107}}, 120401 (2011).

\bibitem{huangPRL}
B. Huang, L. A. Sidorenkov, R. Grimm, and J. M. Hutson,
Observation of the Second Triatomic Resonance in Efimov's Scenario,
Phys. Rev. Lett. {\bf{112}}, 190401 (2014).

\bibitem{efimov_helium}
M. Kunitski, S. Zeller, J. Voigtsberger, A. Kalinin, L. P. H. Schmidt, M. Sch\"offler, A. Czasch, W. Sch\"ollkopf, R. E. Grisenti, 
T. Jahnke, D. Blume, R. D\"orner,
Observation of the Efimov state of the helium trimer, 
Science {\bf{348}}, 551 (2015).

\bibitem{efimov_radio1}
T. Lompe, T. B. Ottenstein, F. Serwane, A. N. Wenz, G Z\"urn, S. Jochim,
Radio-Frequency Association of Efimov Trimers,
Science {\bf{330}}, 940 (2010).

\bibitem{efimov_radio2}
S. Nakajima, M. Horikoshi, T. Mukaiyama, P. Naidon, and M. Ueda,
Measurement of an Efimov Trimer Binding Energy in a Three-Component Mixture of $^6\text{Li}$,
Phys. Rev. Lett. {\bf{106}}, 143201 (2011).

\bibitem{GreenwoodPRD1973}
R. D. Amado and F. C. Greenwood,
There Is No Efimov Effect for Four or More Particles,
Phys. Rev. D {\bf{7}}, 2517 (1973).

\bibitem{SadhanPRD1981}
S. K. Adhikari and A. C. Fonseca,
Four-body Efimov effect in a Born-Oppenheimer model,
Phys. Rev. D {\bf{24}}, 416 (1981).

\bibitem{PlatterPRA2004}
 L. Platter, H.-W. Hammer, and U.-G. Mei{\ss}ner, 
Four-boson system with short-range interactions, 
Phys. Rev. A {\bf{70}}, 052101 (2004).

\bibitem{StecherNatPhys2009}
J. von Stecher, J. P. D'Incao, and C. H. Greene, 
Signatures of universal four-body phenomena and their relation to the Efimov effect, 
Nat. Phys. {\bf{5}}, 417 (2009).


\bibitem{StecherJPhysB2010}
J. von Stecher, 
Weakly bound cluster states of Efimov character,
J. Phys. B {\bf{43}}, 101002 (2010).

\bibitem{NicholsonPRL2012}
A. N. Nicholson, 
$N$-Body Efimov States from Two-Particle Noise, 
Phys. Rev. Lett. {\bf{109}}, 073003 (2012).


\bibitem{GattobigioPRA2014}
M. Gattobigio and A. Kievsky, 
Universality and scaling in the $N$-body sector of Efimov physics, 
Phys. Rev. A {\bf{90}}, 012502 (2014).

\bibitem{YanPRA2015}
Y. Yan and D. Blume,
Energy and structural properties of $N$-boson clusters attached to three-body Efimov states: 
Two-body zero-range interactions and the role of the three-body regulator,
Phys. Rev. A {\bf{92}}, 033626 (2015).

\bibitem{naidon2016}
P. Naidon and S. Endo,
Efimov Physics: a review,
Rep. Prog. Phys. {\bf{80}}, 056001 (2017).

\bibitem{ChrisRMP2017}
C. H. Greene, P. Giannakeas, and J. P\'erez-R\'{\i}os,
Universal few-body physics and cluster formation,
Rev. Mod. Phys. {\bf{89}}, 035006 (2017).


\bibitem{BruunPRL2015}
J. Levinsen, M. M. Parish, and G. M. Bruun,
Impurity in a Bose-Einstein Condensate and the Efimov Effect,
Phys. Rev. Lett. {\bf{115}}, 125302 (2015).

\bibitem{CuiPRL2017}
M. Sun, H. Zhai, and X. Cui,
Visualizing the Efimov Correlation in Bose Polarons,
Phys. Rev. Lett. {\bf{119}}, 013401 (2017).

\bibitem{TaylorPRL2017}
M. J. Gullans, S. Diehl, S. T. Rittenhouse, B. P. Ruzic, J. P. D'Incao, P. Julienne, A. V. Gorshkov, and J. M. Taylor,
Efimov States of Strongly Interacting Photons,
Phys. Rev. Lett. {\bf{119}}, 233601 (2017).

\bibitem{DIncaoPRL2018}
V. E. Colussi, J. P. Corson, and J. P. D'Incao,
Dynamics of Three-Body Correlations in Quenched Unitary Bose Gases,
Phys. Rev. Lett. {\bf{120}}, 100401 (2018).

\bibitem{ParishPRX2018}
S. M. Yoshida, S. Endo, J. Levinsen, and M. M. Parish,
Universality of an Impurity in a Bose-Einstein Condensate,
Phys. Rev. X {\bf{8}}, 011024 (2018). 

\bibitem{CuiPRA2019}
M. Sun and X. Cui,
Efimov physics in the presence of a Fermi sea,
Phys. Rev. A {\bf{99}}, 060701(R) (2019).

\bibitem{JonsellPRL2002}
S. Jonsell, H. Heiselberg, and C. J. Pethick,
Universal Behavior of the Energy of Trapped Few-Boson Systems with Large Scattering Length,
Phys. Rev. Lett. {\bf{89}}, 250401 (2002).

\bibitem{KohlerFoundPhysics2016}
M. Stoll and T. K\"ohler,
Production of three-body Efimov molecules in an optical lattice,
Phys. Rev. A {\bf{72}}, 022714 (2005).

\bibitem{CastinPRL2006}
F. Werner and Y. Castin, 
Unitary Quantum Three-Body Problem in a Harmonic Trap,
Phys. Rev. Lett. {\bf{97}}, 150401 (2006).


\bibitem{KokkelmansFewBody2011}
J. Portegies and S. Kokkelmans,
Efimov Trimers in a Harmonic Potential,
Few-Body Syst. {\bf{51}}, 219 (2011).

\bibitem{BlumeRep}
D. Blume, 
Few-body physics with ultracold atomic and molecular systems in traps,
Rep. Prog. Phys. {\bf{75}}, 4 (2012).

\bibitem{TolleJPhysG2013}
S. T\"olle, H.-W. Hammer and B. Ch. Metsch,
Convergence properties of the effective theory for trapped bosons,
J. Phys. G: Nuclear and Particle Physics {\bf{40}}, 055004 (2013).

\bibitem{ParishPRX2014}
J. Levinsen, P. Massignan, and M. M. Parish,
Efimov Trimers under Strong Confinement,
Phys. Rev. X {\bf{4}}, 031020 (2014).

\bibitem{shi2014}
Z.-Y. Shi, X. Cui, and H. Zhai,
Universal Trimers Induced by Spin-Orbit Coupling in Ultracold Fermi Gases,
Phys. Rev. Lett. {\bf{112}}, 013201 (2014).

\bibitem{cui2014}
X. Cui and W. Yi,
Universal Borromean Binding in Spin-Orbit-Coupled Ultracold Fermi Gases,
Phys. Rev. X {\bf{4}}, 031026 (2014). 

\bibitem{shi2015}
Z.-Y. Shi, H. Zhai, and X. Cui,
Efimov physics and universal trimers in spin-orbit-coupled ultracold atomic mixtures,
Phys. Rev. A {\bf{91}}, 023618 (2015).

\bibitem{GuanPRX2018}
Q. Guan and D. Blume, 
Three-Boson Spectrum in the Presence of 1D Spin-Orbit Coupling: Efimov’s Generalized Radial Scaling Law,
Phys. Rev. X {\bf{8}}, 021057 (2018).


\bibitem{bound_shenoy}
J. P. Vyasanakere and V. B. Shenoy,
Bound states of two spin-1/2 fermions in a synthetic non-Abelian gauge field,
Phys. Rev. B {\bf{83}}, 094515 (2011).

\bibitem{crossover_shenoy}
J. P. Vyasanakere, S. Zhang, and V. B. Shenoy,
BCS-BEC crossover induced by a synthetic non-Abelian gauge field,
Phys. Rev. B {\bf{84}}, 014512 (2011).

\bibitem{rashbon_shenoy}
J. P. Vyasanakere and V. B. Shenoy,
Rashbons: properties and their significance,
New J. Phys. {\bf{14}}, 043041 (2012).

\bibitem{zhenhua2012}
Z. Yu,
Short-range correlations in dilute atomic Fermi gases with spin-orbit coupling,
Phys. Rev. A {\bf{85}}, 042711 (2012).

\bibitem{xiaoling}
X. Cui, 
Mixed-partial-wave scattering with spin-orbit coupling and validity of pseudopotentials,
Phys. Rev. A {\bf{85}}, 022705 (2012).

\bibitem{ShenoyPRA2013}
J. P. Vyasanakere and V. B. Shenoy,
Flow-enhanced pairing and other unusual effects in Fermi gases in synthetic gauge fields,
Phys. Rev. A {\bf{88}}, 033609 (2013).

\bibitem{PuPRA2013}
L. Dong, L. Jiang, H. Hu, and H. Pu,
Finite-momentum dimer bound state in a spin-orbit-coupled Fermi gas,
Phys. Rev. A {\bf{87}}, 043616 (2013).

\bibitem{zhenhua}
Y. Wu and Z. Yu,
Short-range asymptotic behavior of the wave functions of interacting spin-1/2 fermionic atoms with spin-orbit coupling: A model study,
Phys. Rev. A {\bf{87}}, 032703 (2013).

\bibitem{qingze}
Q. Guan and D. Blume,
Scattering framework for two particles with isotropic spin-orbit coupling applicable to all energies,
Phys. Rev. A {\bf{94}}, 022706 (2016).

\bibitem{xiaoling2017}
X. Cui, 
Multichannel molecular state and rectified short-range boundary condition for spin-orbit-coupled ultracold fermions near p-wave resonances,
Phys. Rev. A {\bf{95}}, 030701(R) (2017).

\bibitem{wang2015}
S.-J. Wang and C. H. Greene,
General formalism for ultracold scattering with isotropic spin-orbit coupling,
Phys. Rev. A {\bf{91}}, 022706 (2015).

\bibitem{li_boson}
R. Li and L. Yin,
Pair condensation in a dilute Bose gas with Rashba spin-orbit coupling,
New J. Phys. {\bf{16}}, 053013 (2014).

\bibitem{xu_boson}
Z. Xu, Z. Yu, and S. Zhang,
Evidence for correlated states in a cluster of bosons with Rashba spin-orbit coupling,
New J. Phys. {\bf{18}}, 025002 (2016).

\bibitem{luo_boson}
D. Luo and L. Yin,
BCS-pairing state of a dilute Bose gas with spin-orbit coupling,
Phys. Rev. A {\bf{96}}, 013609 (2017).

\bibitem{YinPRA2018}
Q. Gu and L. Yin,
Spin-orbit-coupling–induced resonance in an ultracold Bose gas,
Phys. Rev. A {\bf{98}}, 013617 (2018).

\bibitem{SpielmanNature2011}
Y.-J. Lin, K. Jim\'enez-Garc{\'\i}a, and I. B. Spielman, 
Spin-orbit-coupled Bose-Einstein condensates, 
Nature (London) {\bf{471}}, 83 (2011).

\bibitem{ECGbook}
Y. Suzuki and K. Varga,
{\it{Stochastic Variational Approach to Quantum-Mechanical Few-Body Problems}}, 
1st ed. (Springer-Verlag Berlin Heidelberg, 1998).


\bibitem{ECGrmp}
J. Mitroy, S. Bubin, W. Horiuchi, Y. Suzuki, L. Adamowicz, W. Cencek, K. Szalewicz, J. Komasa, D. Blume, and K. Varga,
Theory and application of explicitly correlated Gaussians,
Rev. Mod. Phys. {\bf{85}}, 693 (2013).

\bibitem{guan_thesis}
Q. Guan,
One-, two-, and three-body systems with spin-orbit coupling,
Ph.D. thesis, Washington State University (2017).

\bibitem{hui_review}
H. Zhai,
Degenerate quantum gases with spin-orbit coupling: a review,
Rep. Prog. Phys. {\bf{78}}, 026001 (2015).


\bibitem{FeymanPhysRev1939}
R. Feynman,
Forces in Molecules,
Phys. Rev. {\bf{56}}, 340 (1939).


\bibitem{GattobigioPRA2013}
A. Kievsky and M. Gattobigio,
Universal nature and finite-range corrections in elastic atom-dimer scattering below the dimer breakup threshold,
Phys. Rev. A {\bf{87}}, 052719 (2013).

\bibitem{ZhaiPRL2011}
Z.-Q. Yu and H. Zhai,
Spin-Orbit Coupled Fermi Gases across a Feshbach Resonance,
Phys. Rev. Lett. {\bf{107}}, 195305 (2011).

\bibitem{ZhaiRev2012}
H. Zhai,
Spin-Orbit Coupled Quantum Gases,
Int. J. Mod. Phys. B {\bf{26}}, 1230001 (2012).

\bibitem{JiangPRA2012}
S.-G. Peng, X.-J. Liu, H. Hui, and K. Jiang,
Momentum-resolved radio-frequency spectroscopy of a spin-orbit-coupled atomic Fermi gas near a Feshbach resonance in harmonic traps,
Phys. Rev. A {\bf{86}}, 063610 (2012).

\bibitem{HuPRA2013}
X.-J. Liu and H. Hu,
Inhomogeneous Fulde-Ferrell superfluidity in spin-orbit-coupled atomic Fermi gases,
Phys. Rev. A {\bf{87}}, 051608(R) (2013).

\bibitem{HuPRA2013_2}
X.-J. Liu and H. Hu,
Topological Fulde-Ferrell superfluid in spin-orbit-coupled atomic Fermi gases,
Phys. Rev. A {\bf{88}}, 023622 (2013).

\bibitem{PuPRA2014}
L. Jiang, E. Tiesinga, X.-J. Liu, H. Hu, and H. Pu,
Spin-orbit-coupled topological Fulde-Ferrell states of fermions in a harmonic trap,
Phys. Rev. A {\bf{90}}, 053606 (2014).

\bibitem{EiteRMP2010}
C. Chin, R. Grimm, P. Julienne, and E. Tiesinga,
Feshbach resonances in ultracold gases,
Rev. Mod. Phys. {\bf{82}}, 1225 (2010).

\bibitem{galitski_review2013}
V. Galitski and I. B. Spielman,
Spin-orbit coupling in quantum gases,
Nature (London) {\bf{494}}, 49 (2013).

\bibitem{goldman2014}
N. Goldman, G. Juzeli{\={u}}nas, P. \"Ohberg, and I. B. Spielman,
Light-induced gauge fields for ultracold atoms,
Rep. Prog. Phys. {\bf{77}}, 126401 (2014).

\end{thebibliography}
\end{document}